%% file: main.tex
\documentclass[acmsmall]{acmart}
\usepackage{colortbl}
\usepackage{arydshln}
\usepackage{multirow}
\usepackage{multicol}
 \usepackage{array}
\usepackage{bm}
\usepackage{booktabs} 
\usepackage{amsmath}
\usepackage{amsfonts}
\usepackage{graphicx}
\usepackage{tabularx}
\usepackage{textcomp}
\usepackage{xcolor}
\usepackage{listings}
\usepackage{multirow}
\usepackage{subfig}
\usepackage{algorithm}
\usepackage[noend]{algpseudocode}
\usepackage{graphicx}
\usepackage{bbding}

\AtBeginDocument{%
  \providecommand\BibTeX{{%
    \normalfont B\kern-0.5em{\scshape i\kern-0.25em b}\kern-0.8em\TeX}}}

\setcopyright{acmcopyright}
\copyrightyear{2021}
\acmYear{2021}
\acmDOI{10.1145/1122445.1122456}

\acmJournal{JACM}
\acmVolume{37}
\acmNumber{4}
\acmArticle{111}
\acmMonth{8}

\begin{document}

\title{Time-aware Path Reasoning on Knowledge Graph for Recommendation}

\newcommand\norm[1]{\left\lVert#1\right\rVert}
\newcommand{\Lapl}{\mathbf{\mathop{\mathcal{L}}}}
\newcommand{\lapl}{\mathcal{L}}
\newcommand{\Trans}[1]{{#1}^{\top}}
\newcommand{\Trace}[1]{tr\left({#1}\right)}
\newcommand{\Bracs}[1]{\left({#1}\right)}
\newcommand{\Mat}[1]{\textbf{#1}}
\newcommand{\MatS}[3]{\mathbf{#1}^{#2}_{#3}}
\newcommand{\Space}[1]{\mathbb{#1}}
\newcommand{\Set}[1]{\mathcal{#1}}
\newcommand{\vectornorm}[1]{\left|\left|#1\right|\right|}
\newcommand{\bpi}{\boldsymbol{\pi}}
\newcommand{\BlockMat}[2]{\left[\begin{matrix}#1\\#2\end{matrix}\right]}
\newcommand{\BlockMatSquare}[4]{\left[\begin{matrix}#1 & #2\\#3 & #4\end{matrix}\right]}
\newcommand{\ParaTitle}[1]{\noindent\textbf{#1}}
\newcommand\at[2]{\left.#1\right|_{#2}}

\newcommand{\ie}{{\em i.e.}}
\newcommand{\eg}{{\em e.g.}}
\newcommand{\etal}{{\em et al.}}
\newcommand{\etc}{{\em etc.}}
\newcommand{\tkgrec}{TPRec}



\author{Yuyue Zhao}
\email{yyzha0@mail.ustc.edu.cn}
\affiliation{%
  \institution{University of Science and Technology of China}
  \country{China}
  }

\author{Xiang Wang}
\authornote{Corresponding Author.}
\email{xiangwang1223@gmail.com}
\affiliation{%
  \institution{University of Science and Technology of China}
  \country{China}
  }
  
\author{Jiawei Chen}
\email{cjwustc@ustc.edu.cn}
\affiliation{%
  \institution{University of Science and Technology of China}
  \country{China}
  }
  
  \author{Yashen Wang}
\email{yswang@bit.edu.cn}
\affiliation{%
  \institution{National Engineering Laboratory 
        for Risk Perception and Prevention (RPP)}
  \country{China}
  }

 \author{Wei Tang}
\email{weitang@mail.ustc.edu.cn}
\affiliation{%
  \institution{University of Science and Technology of China}
  \country{China}
  }

\author{Xiangnan He}
\email{xiangnanhe@gmail.com}
\affiliation{%
  \institution{University of Science and Technology of China}
  \streetaddress{96 JinZhai Road}
  \city{Hefei}
  \country{China}
}

\author{Haiyong Xie}
\authornote{Corresponding Author.}
\email{haiyong.xie@ieee.org}
\affiliation{
  \institution{Key Laboratory of Cyberculture Content Cognition and Detection, Ministry of Culture and Tourism, }
  \institution{Advanced Innovation Center for Human Brain Protection, Capital Medical University}
  \country{China}
  \streetaddress{The authors from the University of Science and Technology of China are affliated with the CCCD Key Lab of MCT}
  }

\renewcommand{\shortauthors}{Zhao, et al.}


\input{0-abstract}



\maketitle

\input{1-introduction}

\input{2-ProblemFormulation}

\input{3-Method}

\input{4-Experiment}

\input{5-RelatedWork}

\input{6-Conclusion}

\input{acknow}


\bibliographystyle{ACM-Reference-Format}
\bibliography{sigproc}

\end{document}

%% file: 0-abstract.tex
\begin{abstract}

Reasoning on knowledge graph (KG) has been studied for explainable recommendation due to its ability of providing explicit explanations.
    However, current KG-based explainable recommendation methods unfortunately
    ignore the temporal information (such as purchase time,
    recommend time, \etc), which may result in unsuitable explanations.
%
In this work, we propose a novel  {\it Time-aware Path reasoning 
for Recommendation} (\tkgrec\ for short) method, which leverages the potential 
of temporal information to offer better recommendation with plausible
    explanations.
First, we present an efficient time-aware interaction relation 
extraction component
    to construct collaborative knowledge graph with time-aware interactions 
    (TCKG for short), and then 
    introduce
    a novel time-aware path reasoning  method for recommendation.  We conduct
    extensive experiments on three real-world datasets. The results demonstrate that the proposed \tkgrec\ could successfully
    employ TCKG to achieve substantial gains and improve the quality of 
    explainable recommendation. 

\end{abstract}

%% file: 1-introduction.tex
\section{Introduction}


Recently, knowledge graphs (KGs) have been widely used in 
recommender systems
due to their rich structured knowledge
 (\eg, \cite{ripplenet, pgpr, CKE, DKN, KSR, shine, 
ADAC, wu2021disenkgat, zeng2022shadewatcher,
RNNexpl}). A KG is a type of directed heterogeneous graph in which nodes 
represent
real-world \textit{entities} and edges represent their \textit{relations}.
%
Reasoning over KGs \cite{pgpr, ADAC} for recommendation can not 
only infer user preferences more accurately, but also
offer each recommended item with a multi-hop path. Such 
knowledge-aware paths allow us to explain why an item is recommended, so as to increase users' satisfaction and trust \cite{surveyExp, 
effectiveRec}.
%
Existing methods adopt reinforcement learning agents to automatically mine and reason over such paths.
The core lies in  using 
\textbf{"What items will users purchase?"}
 to guide the 
reasoning process towards the target items.



\begin{figure}[!t]
\includegraphics[width=0.7\columnwidth]{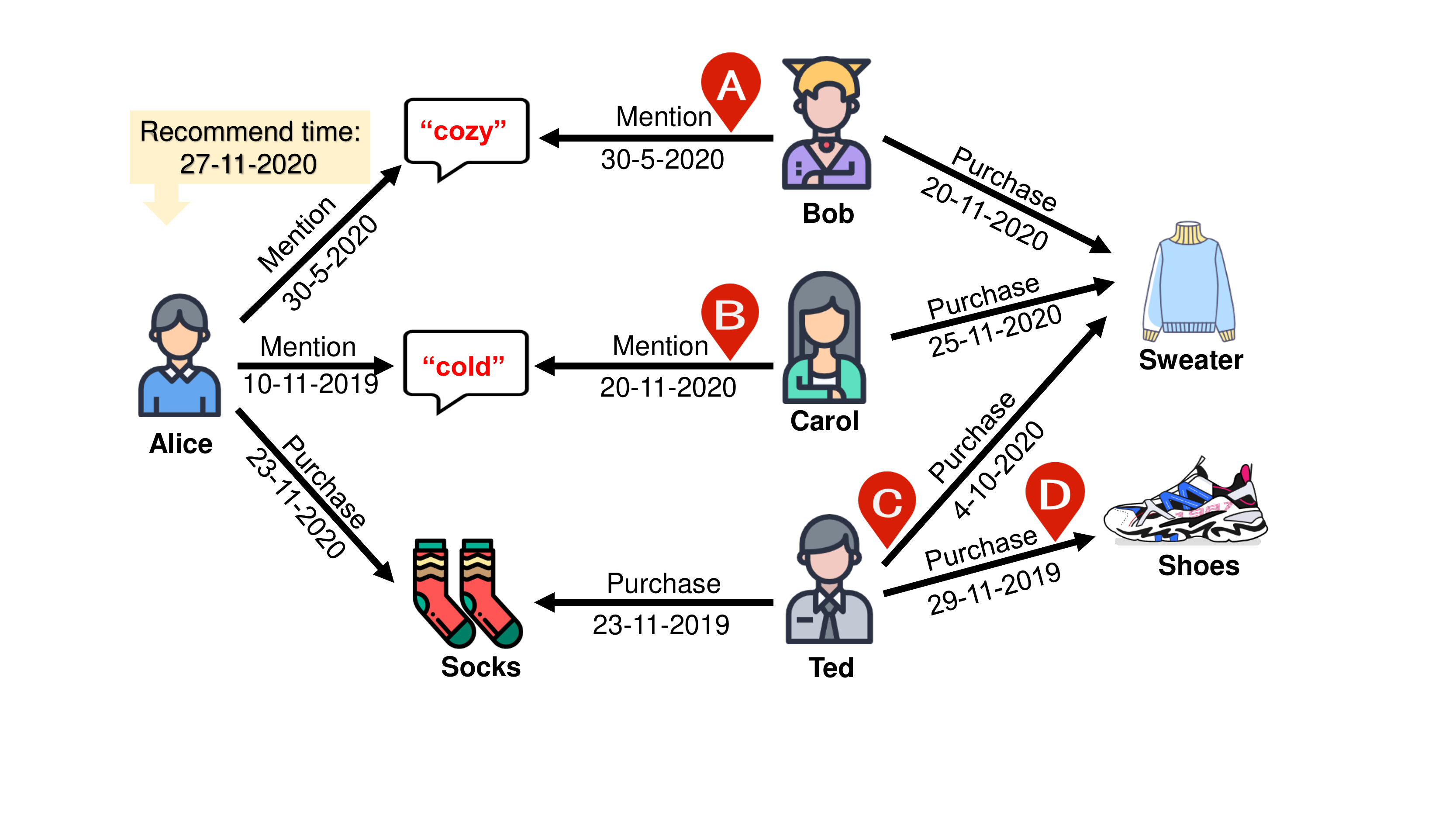}
\caption{Example of how temporal information assists in explainable 
        recommendation.}
\label{intro_exp}
\end{figure}


However, current RL agents mostly treat the user-item interaction as the static 
relation, and ignore the time-aware information of interactions (\eg, purchase 
time).
Hence, they fall short in modeling the temporal patterns of user behaviors, 
possibly offering unreliable reasoning paths and recommendations.
Considering the example in Figure
\ref{intro_exp}, where A, B, C and D marked in red indicate different reasoning
paths. 
Typical KG reasoning methods (\eg, \cite{pgpr, ADAC}) may recommend
$Sweater$ to $Alice$, which can be explained by the path A existing in the
corresponding KG, namely, $Alice \stackrel{Mention}{\longrightarrow}\text{cozy}
\stackrel{Mentioned\_by}{\longrightarrow}Bob
\stackrel{Purchase}{\longrightarrow}Sweater$. 
However, with the help of temporal information, it can be shown that
the reasoning paths B and D are better than  A and C. 
First, in explaining the same recommendation $Sweater$, path B is more reasonable than A since its relation time (10-11-2019) is
in the season of winter, which is consistent with the recommendation time
(27-11-2020).  This implies the periodic patterns of user behaviors. 
Second, path D is better than C since the purchase time 29-11-2019 and
recommendation time 27-11-2020 are both Black Friday while 4-10-2020 is closer to 27-11-2020 only in linear time.
People tend to have similar behaviors in shopping festivals.
In this sense, we consider Path D a better recommendation.
Therefore, a time-aware KG reasoning method is more 
likely to generate better recommendation results.

%
\begin{figure}[!t]
    \begin{minipage}[b]{\columnwidth}
        \centering
        \subfloat[Clothing]{
            \centering
            \includegraphics[width=0.33\columnwidth]{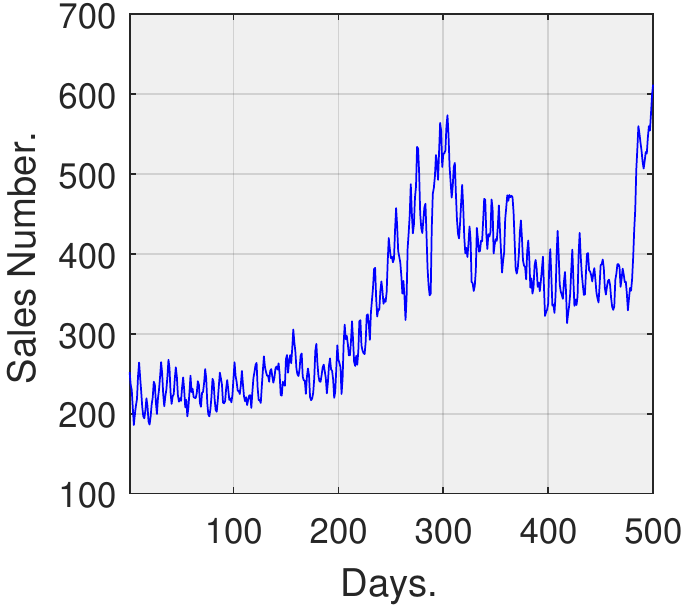}
            \label{day_Cloth}
        }
        \subfloat[Cell Phones.]{
            \centering
            \includegraphics[width=0.33\columnwidth]{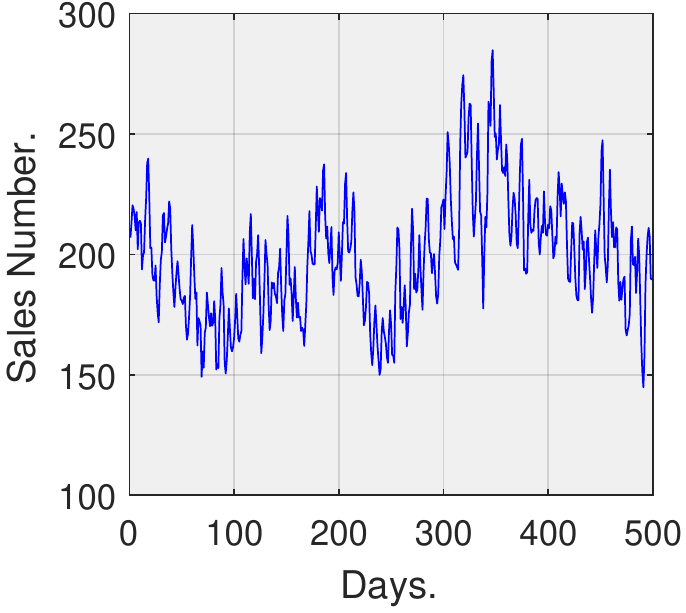}
            \label{day_Cell}
        }
        \subfloat[Beauty]{
            \centering
            \includegraphics[width=0.33\columnwidth]{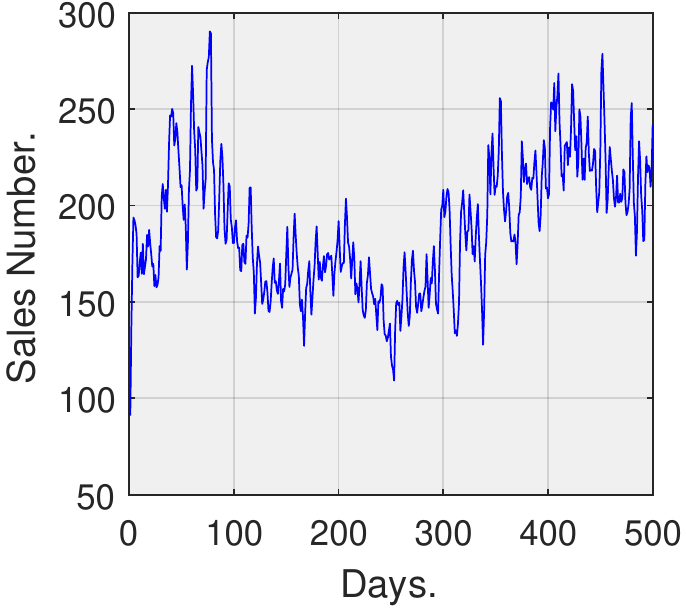}
            \label{day_Beauty}
        }
        \caption{Sales behavior in three Amazon datasets.} 
        \label{day_cmp}
    \end{minipage}
    \vspace{-0.15cm}
\end{figure}

 While some KG-based recommendation methods have modeled temporal information~\cite{kerl, KSR, huang2019explainable}, they mainly use KG data to enhance representation rather 
reasoning over the KG. Their focus is on modeling the interaction sequence for sequential recommendation, rather than the more general temporal patterns as exampled above. It remains underexplored how to model the rich patterns (\eg, seasons, shopping festivals, \etc) that can be inferred from timestamps.  


In this work, we
frame user-item interactions as time-aware relations. In the example of Figure \ref{intro_exp}, if the reasoning agent is guided by  
\textbf{"What items will users purchase in winter and festival?"}
, then 
\underline{Path B} and \underline{Path D}  can be easily inferred.
Nonetheless, it is non-trivial due to 
the following challenges: 
%
\textbf{(i) A huge number of timestamps.}
Figure \ref{day_cmp} is based on the Amazon datasets~\cite{HM+2016},  which 
demonstrate user purchasing behaviors, as well as 
the corresponding timestamps over a 500-day period.
As the total number of timestamps is numerous, they cannot be directly 
considered as relation types.
%
\textbf{(ii) 	
Ambiguous definition of temporal distance.}
Those timestamps 
do not always suggest similar user behaviors.
As illustrated in Figure \ref{day_cmp}, 
the sales of items in the same season tend to be similar in different years,
rather than in different seasons of the same year. 
%
%
%
\textbf{(iii) Different information sources.}
In recommender systems, the interactions between users and items continuously stream in. But the external KG is relatively stable and seldom updated,
%
%
which means we couldn't treat them equally 
when performing time-aware path reasoning.

To address the above challenges, 
we integrate the time-aware interactions into KG as a 
heterogeneous information graph (TCKG for short).
%
We propose a novel
{\it \textbf{T}ime-aware \textbf{P}ath reasoning for \textbf{Rec}ommendation}
(\tkgrec\ for short) method. 
Our method consists of three components.
The first one is the time-aware interaction relation extraction 
component.
%
We apply time series clustering on 
the interaction timestamps
and replace the original interaction relations with clustered relations.
By doing this, we could control the number of relation types, and make the 
extracted relations sensitive to time.
More specifically, the timestamp's temporal features consist of statistical and 
structural
discriminative properties, which enables us to measure temporal similarity and 
behavioral pattern similarity.  
In the second component, we adopt translational distance learning to 
encode entities and relations in TCKG to obtain
time-aware semantic representation.
The third component is the key time-aware path reasoning component.
In this component, we shape a personalized time-aware reward 
based on each user's interaction history and recommend time,
which is used to guide the reinforcement learning (RL) reasoner to explore more
precisely.
%
%
%

%

In summary, the contributions are as follows:

\begin{itemize}
\item We design a TCKG to properly integrate the temporal information into 
KG-based recommendation.
To the best of our knowledge, it is the first work to 
introduce time-aware path reasoning
for explainable recommendation.

%

\item We propose a personalized time-aware reward to guide the RL-based 
reasoning agent,
which makes the time-aware reasoning achieve a
more accurate and interpretable  recommendation.

\item We conduct experiments on three real-world datasets, and the results 
demonstrate the effectiveness of \tkgrec\ over several state-of-the-art 
baselines. The codes are available at https://github.com/Go0day/TPRec.

\end{itemize}


%% file: 2-ProblemFormulation.tex
\section{Problem Formulation}\label{sec2}

%

\begin{table*}[t!] 
    \centering
    \caption{Notation and Definitions.}
\begin{tabular}{cc} 
\toprule
\textbf{Notations} & \textbf{Annotation} \\ \hline
$\mathcal{U}$                  & User set                \\
$\mathcal{V}$                  & Global item set                \\ 
$\mathcal{V}_\mathcal{U}$      & The observed interactions between users and items  \\
$\mathcal{V}_\mathcal{U}^T$     & The clustered time-aware interaction set, with $L$ categories \\
$\mathcal{G}_T$                 & Time-aware Collaborative Knowledge Graph \\
$\mathcal{E}^{\prime}$          & The entities in Time-aware Collaborative Knowledge Graph \\
$\mathcal{R}_T$                 & The relations in Time-aware Collaborative Knowledge Graph \\
$T$                             & The interaction timestamp set \\
$\hat{T}$                       & The recommend time \\
$\hat{\mathcal{V}}$             & The recommended item set \\
$\tau_{u, \hat{\mathcal{V}}}$   & The reasoning path set for user $u$ to  item set $\hat{\mathcal{V}}$ \\
$\mathcal{T}$                & Temporal feature space, including statistical features and structural features \\
$r_{kg}$                        & The  external knowledge graph relation set \\
$r_{interact}$                  & The time-aware interaction relation set \\
$\Mat{r}$                       & The embedding of relations            \\
$\Mat{e}_h, \Mat{e}_t$          & The embedding of entities                \\
$\Mat{W}_{t},  \Mat{W}_{\hat{T}}$   & The weight distribution of the Gaussian models at timestamp $t$,\ $\hat{T}$ \\
$\Mat{W}_{h_u}$                 & The time cluster weight distribution of user $u$'s interaction history \\
$k, k^{\prime}, K$                  & The $k$-th reasoning step, state history length $k^{\prime}$ and terminal step \\
$\Mat{s}_k$                     & The state at step $k$ \\
$a_k, \Set{A}_k$                     & The action and action space at step $k$ \\
$g_R(\hat{v} \mid u)$           &  The time-aware based reward for recommended item $\hat{v}$ for user $u$ \\
$\pi\left(\cdot \mid \mathbf{s}_k, \mathcal{A}_{k}(u)\right)$ & The policy network at step $k$ \\
$\hat{c}(\mathbf{s}_k)$         & The value network at step $k$ \\ \bottomrule
\end{tabular}
\label{notation_table}
\end{table*}

\noindent\textbf{Notions:} 
We consider the time-aware path reasoning for KG-enhanced recommendation.
Let denote the user set, the item set, and the observed interactions as $\mathcal{U}$, $\mathcal{V}$, ${\mathcal{V}}_{\mathcal{U}}$, respectively.
Moreover, there exists an external KG storing item properties. Generally, the KG is represented as $\mathcal{G}=\{(h, r, t)  \mid h, t \in \mathcal{E}, r \in \mathcal{R}\}$, where $\mathcal{E}$ represents the set of real-world entities and $\mathcal{R}$ is the set of relation.  
Each triplet $(h,r,t)$ delineates that there exist a relationship $r$ from head entity $h$ to tail entity $t$.
Previous works \cite{multi_modal,kgat} combine the user behaviors and item knowledge as a unified relational graph, termed Collaborative Knowledge Graph (CKG), which is defined as 
$\mathcal{G}_R=\{(h, r, t) \mid h, t \in \mathcal{E}^{\prime}$, $ r \in \mathcal{R}^{\prime}\}$,  where $ \mathcal{E}^{\prime}=\mathcal{E} \cup \mathcal{U}$ and $ \mathcal{R}^{\prime}=\mathcal{R} \cup{\mathcal{V}}_{\mathcal{U}}$. The notations of this work are summarized in Table \ref{notation_table}.
	 
\vspace{5pt}
\noindent\textbf{Reasoning Environment:}
Considering the time-aware interactions instead, we further obtain a Time-aware Collaborative Knowledge Graph (TCKG), which is the environment of time-aware path reasoning.
Specifically, the interaction timestamps are grouped into $L$ groups after time-series clustering, and the interaction relation ${\mathcal{V}}_{\mathcal{U}}$ is extended to $\{\mathcal{V}_{\mathcal{U}}^1, 
\mathcal{V}_{\mathcal{U}}^2, \cdots, \mathcal{V}_{\mathcal{U}}^{L}\}$.
Hence, the TCKG is formulated as $\mathcal{G}_T=\{(h, r, t) \mid 
h, t \in \mathcal{E}^{\prime}, r \in \mathcal{R}_{T}\}$, where 
$ \mathcal{R}_{T}=\mathcal{R} \cup \{ \mathcal{V}_{\mathcal{U}}^1, 
\mathcal{V}_{\mathcal{U}}^2, \cdots, \mathcal{V}_{\mathcal{U}}^{L} \}$.

\vspace{5pt}
\noindent\textbf{Task Description:}
We present the task of time-aware path reasoning for recommendation (TPRec) as follows:
\begin{itemize}
\item \textbf{Input:} Time-aware collaborative knowledge graph $\mathcal{G}_T$;

\item \textbf{Output:} Given a user $u$ and a timestamp $\hat{T}$, a recommender model results in an item set $\hat{\mathcal{V}} \subseteq 
\mathcal{V} $ with the corresponding explainable reasoning paths $\tau_{u, 
\hat{\mathcal{V}}}$.  
In particular, for an item $\hat{v} \in \hat{\mathcal{V}}$, 
$\tau_{u, \hat{v}}$ is a multi-hop path, which connects user $u$ with item $\hat{v}$ to explain why $\hat{v}$ is suitable for $u$: 
$\tau_{u, \hat{v}}=[u 
\stackrel{r_{1}}{\longrightarrow} e_{1} \stackrel{r_{2}}{\longrightarrow} 
\ldots \stackrel{r_{k-1}}{\longrightarrow} e_{k-1} 
\stackrel{r_{k}}{\longrightarrow} \hat{v}]$, where $e_{i-1} 
\stackrel{r_{i}}{\longrightarrow} e_{i}$ represents $(e_{i-1}, r_i, e_i) \in 
\mathcal{G}_T, i \in [k]$.

\end{itemize} 


%% file: 3-Method.tex
\section{Methodology}
\label{sec:method}


%
In this section, we introduce the proposed \tkgrec\ in detail.
The overall framework of time-aware path reasoning is shown in Figure 
\ref{pipeline}.
Our approach is able to effectively fuse temporal information into knowledge 
graph and leverage a RL framework for recommendation.
In what follows, we start with 
\textbf{Time-aware Interaction Relation Extraction} 
to construct TCKG, and use \textbf{Time-aware representation learning} 
to generate entities and relations representations.
Then we present \textbf{Time-aware Path Reasoning} and \textbf{Model Prediction} to reason over TCKG for recommendation.


\subsection{Time-aware Interaction Relation Extraction}\label{tire}
Clearly, how to deal with the dynamic characteristics of user-item interactions is of crucial importance in our task.
However, simply treating the interaction timestamps as auxiliary features might limit the potential of time-aware reasoning, due to the following challenges:
(1) As the timestamps are usually numerous \cite{datasetAmazon}, directly attaching them with user behaviors might cause meaningless features, such as purchase\_20150101 and purchase\_20150102, thus enlarging the feature space dramatically and losing the periodic semantics of timestamps.
(2) The interaction graph and knowledge graph come from heterogeneous sources, and thus are hardly updated synchronously --- that is, the interactions are usually dynamic, while the item knowledge (\eg, category, brand) is fixed over different timestamps.

\begin{figure}[tb]
\includegraphics[width=0.85\columnwidth]{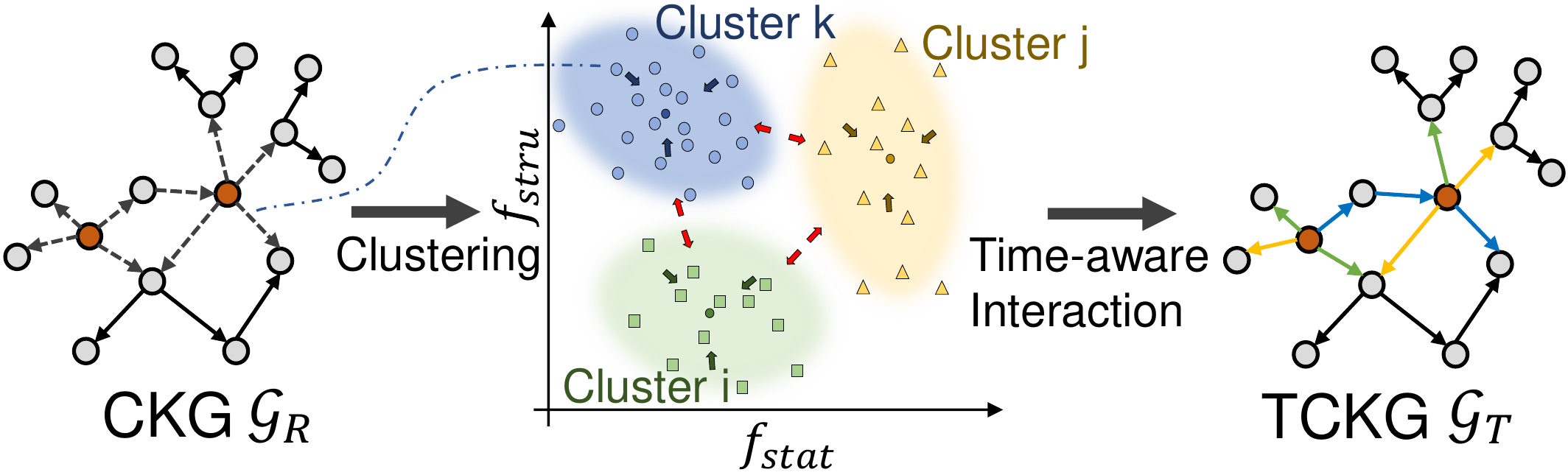}
\caption{Process of time-aware interaction relation extraction. Each 
interaction relation's (dashed edge) timestamp will be mapped into temporal 
feature space 
$\mathcal{T}$ and clustered. Different colored edges (\eg, blue, yellow, green) in TCKG represent 
different clustering results. }
\label{clu}
\end{figure}

To solve these challenges, we first consider the temporal and behavioral characteristics of interactions by projecting the timestamps into a temporal feature space $\mathcal{T}$.
It encodes temporal statistical and temporal structural features of interaction timestamps.
We then group the interaction timestamps $T = \{t_1, t_2, \cdots, t_n\}$ into $L$ categories, and then extend the single interaction type into $\mathcal{V}_{\mathcal{U}}^T = \{ \mathcal{V}_{\mathcal{U}}^1, \mathcal{V}_{\mathcal{U}}^2, \cdots, \mathcal{V}_{\mathcal{U}}^{L} \}$.
After that, the temporal user-item graph can be seamlessly integrated with relatively static KG based on the item-entity alignment set, and finally get the TCKG we need.
The extraction process is illustrated in Figure \ref{clu}.

\subsubsection{\textbf{Temporal Statistical Features}}
Before clustering timestamps, we project them into a feature space.
Specifically, the statistical features are considered to capture some time rules like seasonal changes, periodic purchases \cite{liao2005clustering}.
Hence, for a specific timestamp $t$, its statistical feature is formally summarized as:
\begin{equation}
    \boldsymbol{f}_{stat} =year(t)|| season(t) || month(t) || week(t) \cdots,
\end{equation}
where $\boldsymbol{f}_{stat}$ encodes the year,  season, month, week, and other statistical characteristics of $t$ and concatenates them together, where $||$ is the concatenate operator.

\subsubsection{\textbf{Temporal structural features}}
Beyond the statistical features, we also exploit the structural features to delineate the characteristics of user behaviors, which measures the purchase propensity (\eg, shopping festival) at timestamp $t$.
More formally, we define the structural feature of $t$ as:
\begin{equation}
	\boldsymbol{f}_{stru} = z_{gap}'(t) || z_{gap}''(t),
\end{equation}
where $z_{gap}'(t)$ is the first-order structural feature, which measures the trend of interaction number in the current $gap$ time, compared with that in the past $gap$; $gap$ represents the granularity of window period, which can be set as 30 (\ie, month), or 1 (\ie, day); analogously, $z_{gap}''(t)$ denotes the second-order structural feature and models the momentum of interaction trend.
More formally, $z_{gap}'(t)$ and $z_{gap}''(t)$ is defined as:
\begin{align}  \label{z_p}
	 z_{gap}'(t) &= \frac{\sum_{i=t-gap}^{t}z(i) - \sum_{i=t-2gap}^{t-gap}z(i)}{gap},\ & z_{gap}''(t) &= \frac{\sum_{i=t-gap}^{t}z_{gap}'(i) - \sum_{i=t-2gap}^{t-gap}z_{gap}'(i)}{gap},
\end{align}

where $z(i)$ is the interaction number happened on timestamp $i$, and we replace $z(i)$ with $z_{gap}'(i)$ in Eq. \ref{z_p} to calculate $z_{gap}''(t)$.
For those timestamp $t$ has less than $2gap$ historical interactions, their historical information is not enough to calculate $z_{gap}'(i)$.
In that case, we padding their corresponding features with the nearest timestamp's temporal structural feature. The same goes for $z_{gap}''(t)$.
In this work, we set the gap as 90 (\ie, season), 30 (\ie, month), 7 (\ie, week), 1 (\ie, day) and simply concatenate them as temporal structural features.

Thereafter, we combine these two features together as the temporal feature space $\mathcal{T} \in \Space{R}^m$ for timestamp $t$:
\begin{equation}
	\mathcal{T} = \boldsymbol{f}_{stat} \ || \ \boldsymbol{f}_{stru},
\end{equation}
where $m$ is the feature dimension.

\subsubsection{\textbf{TCKG Construction}}
Having mapped the timestamp to the temporal feature space, we now construct the relation of time-aware interaction.
Here we adopt a time-series clustering method, Gaussian Mixture Model (GMM) \cite{gmm}, on the timestamp set $T$, so as to reduce the timestamp numbers and discretize them into group values.
We leave the exploration of other clustering methods, such as K-means \cite{kmeans}, DBSCAN \cite{ester1996density}, and Mean-shift \cite{meanshift}, in future work.

Specifically, we hire $L$ Gaussian models. For the timestamp $t$ with its temporal feature, we can get the probability $w_{l}$ of being generated by the $l$-th Gaussian model.
As such, the interaction relation $r_t$ on timestamp $t$ can be obtained as follows:
\begin{equation}\label{relation_cluster_2}
	r_t \leftarrow \Space{I}_{max(\Mat{W}_{t})}(\Mat{W}_{t}) {\mathcal{V}_{\mathcal{U}}^T} \ ,
\end{equation}
where $\Mat{W}_{t} = [w^{1}_t, w^{2}_t, \cdots, w^{L}_t]^{\mathsf{T}}$ indicates the $t$-specific probability distribution derived from the Gaussian models;
${\mathcal{V}}_{\mathcal{U}}^T = [\mathcal{V}_{\mathcal{U}}^1, \mathcal{V}_{\mathcal{U}}^2, \cdots, \mathcal{V}_{\mathcal{U}}^{L}]$ is the time-aware interaction relation set, where $\mathcal{V}_{\mathcal{U}}^{l}$ is the $l$-th clustered relation.
Equation \eqref{relation_cluster_2} replaces $r_t$ with the most probable relation in ${\mathcal{V}}_{\mathcal{U}}^T$.
The number of time clusters $L$ is determined by the Bayesian Information Criterion (BIC) \cite{bic}, which is adjusted adaptively for different datasets.

Through the above extract process, we are able to project each observed interaction with timestamp into the same time-aware feature space, and divide all timestamps into weighted $L$ clusters by GMM.
The clustering assignments are used to replace the original interaction relations in CKG and finally convert CKG into TCKG.

\begin{figure}[tb]
\includegraphics[width=\columnwidth]{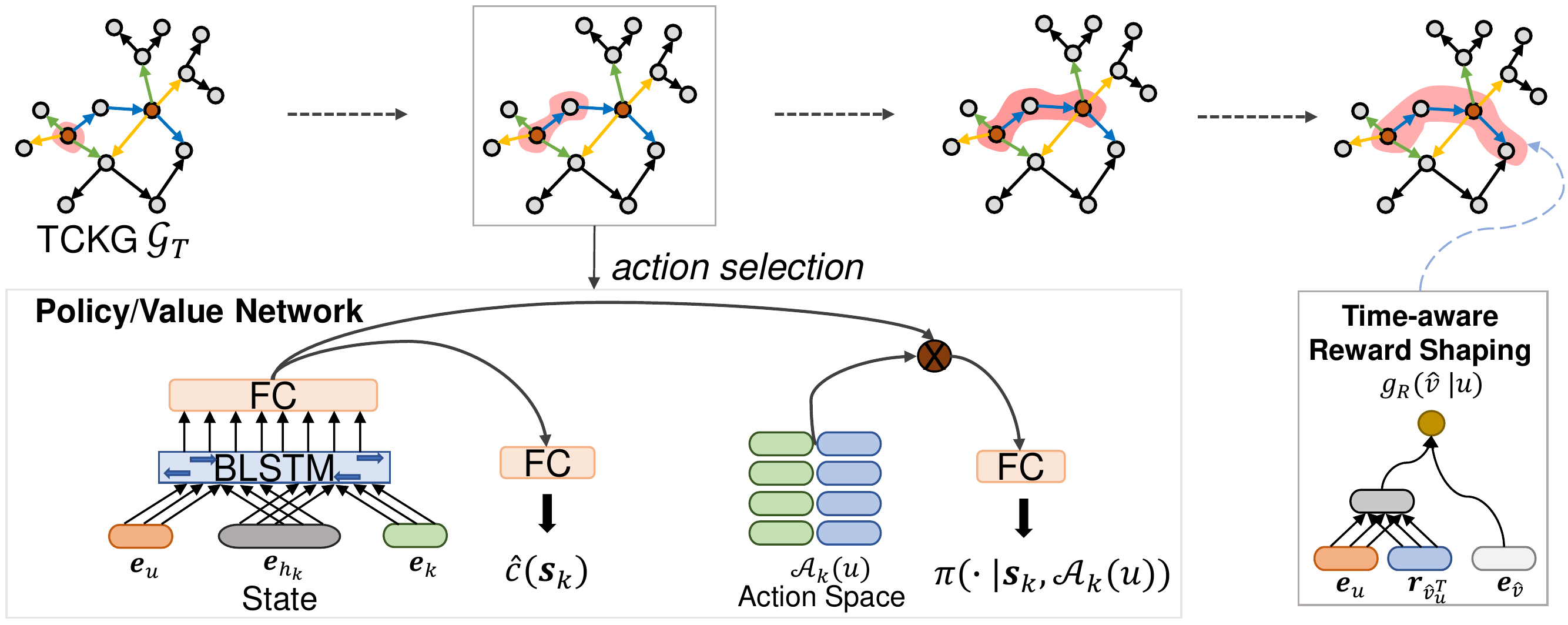}
\caption{ Overall training approach of the proposed \tkgrec\ model.
		For a user $u$ at each time step $k$, as shown in the pink shaded part of TCKG, the agent chooses the next hop based on the temporal constraint, which samples a  
		link $a_k$ according to 
		$\pi\left(\cdot \mid \mathbf{s}_k, \mathcal{A}_{k}(u)\right)$.
		The agent receives a reward $g_R(\hat{v} \mid u)$  when stopped at an item 
		$\hat{v}$. The pink shaded path starts at user $u$ and ends at item $\hat{v}$, showing why our \tkgrec\ model recommends $\hat{v}$ to $u$.
		%
				}
		\label{pipeline}
\end{figure}

\subsection{Time-aware Representation Learning}\label{trl}
Having constructed TCKG, we aim to learn representations for entities and relations by exhibiting the structural and temporal information.
In particular, the KG relations (defined as $r_{kg}$) require the entity representations to meet the structural constraints and maintain the semantics, while the time-aware interaction relations (defined as $r_{interact} \in {\mathcal{V}}_{\mathcal{U}}^T $) encourage the relation affinity within the same categories compared to that of different categories.

Towards this end, we employ the widely-used knowledge graph embedding method, TransE \cite{transE}, on TCKG.
To be more specific, for a given TCKG triplet $(h,r,t)$, it represents head entity $h$, relation $r \in \{r_{kg} \cup r_{interact}\}$ 
, and tail entity $t$ with $d$-dimension embeddings, \ie, $\Mat{e}_{h}$, $\Mat{r}$, $\Mat{e}_{t}$ $\in \mathbb{R}^d$.
These embeddings follow the translational principle: $\Mat{e}_h + \Mat{r} \approx \Mat{e}_t$.
Hence,  for a given triplet $(h, r, t)$, the scoring function is defined as the distance between $\Mat{e}_h + \Mat{r}$ and $\Mat{e}_t$, \ie,
\begin{equation}\label{trans}
	g_{r}({h},{t})=\|\Mat{e}_h+\Mat{r}-
	\Mat{e}_t\|_2^2 .
\end{equation}
A smaller score $g_{r}(h, t)$ suggests that the fact $(h, r, t)$ is more likely to be true, and vice versa. 
We employ the pairwise loss function to train the TCKG embeddings, which accounts for the relative order between the valid triplets and broken ones:
\begin{equation}\label{eqtranse}
    \Lapl_{tckg}=\sum_{\left(h, r, t, t'\right) \in \Set{S}}-\ln\sigma\left(g_{r}({h},{t}') - g_{r}({h},{t})\right)
\end{equation}
where $\Set{S} = \{(h, r, t, t') | (h, r, t) \in \Set{G}_T, (h, r, t') \notin \Set{G}_T\}$, and $ (h, r, t')$ is collected with the tail $t$ replaced by a random entity $t'$; 
$\sigma(\cdot)$ is the sigmoid function. 
This loss serves as a constraint that injects the structural and temporal information into representations, thus preserving the semantics of entities and relations.

\subsection{Time-aware Path Reasoning}\label{tpr}
Having established the embeddings of entities and relations, we feed them into the Time-aware Path Reasoning module. We utilize path reasoning to identify a set of recommended items $\hat{\mathcal{V}}$ for each user $u$ as well as reasoning paths $\{\tau_{u, \hat{v}} | \hat{v} \in  \hat{\mathcal{V}}\}$ for the recommendation.
Nonetheless, existing path reasoning models \cite{pgpr,ADAC} fail to consider temporal information, which might hurt the recommendation performance and result in improper explanations (as illustrated in Figure \ref{intro_exp}). To solve this issue, we introduce temporal information by setting personalized time-aware reward during path reasoning.

\subsubsection{\textbf{Markov Decision Process Formulation}}\label{secMDP}
We follow the previous studies and use the Markov Decision Process (MDP) \cite{reinforcement} in reinforcement learning (RL) to set up the reasoning environment.
The environment can (1) inform the reasoner about its search state and available actions in the TCKG, and (2) reward the reasoner how well the current policy fits the observed user interactions. Overall, the MDP consists of the following components:

\textbf{State.} The initial state is $\Mat{s}_0 = (u, u, \varnothing)$, and the state at step $k$ is $\Mat{s}_k = (u, h_k, e_k)$, where $u \in \Set{U}$ denotes the initial user entity, $e_k$ is the entity which the agent has reached at step $k$, and $h_k$ is the history prior to step $k$. Moreover, to control the model size, we adopt $k'$-step history to encode the combination of all entities and relations in the past $k'$ steps, \ie, $h_k = \{e_{k-k'}, r_{k-k'+1},\cdots,e_{k-1},r_{k}\}$.

\textbf{Action.} For state $\Mat{s}_k$ at step $k$, the reasoner generates an action $a_k = (r_{k+1}, e_{k+1}) \in \mathcal{A}_k$, where $e_{k+1}$ is the next entity to visit, and $r_{t+1}$ is the relation between $e_{k+1}$ and the current entity $e_{k}$. While maintaining the size of the action space based on the largest out-degree may lead to better model efficiency, we also need to control space complexity to reason efficiently.
Therefore, we adopt a pruning function $g_k((r,e_{k+1})|u)$ to reduce the action space on user $u$, as follows:
\begin{equation}\label{actionprun}
\Set{A}_k(u) = \{(r,e_{k+1}) \ |\  \text{rank}(g_k((r,e_{k+1}) \, |\, u)) \leq \epsilon\ \},
\end{equation}
where $(e_k, r, e_{k+1}) \in \mathcal{G}_K$, and $\epsilon$ is the pre-defined action space size. Inspired by \cite{pgpr}, the pruning function is set as
$g_k((r, e_{k+1})|u) = (\boldsymbol{e}_u + \sum_{k''=1}^k\boldsymbol{r}_{k''}) \cdot \boldsymbol{e}_{k+1} + \boldsymbol{b}_{{e}_{k+1}}$, 
where $\cdot$ is the inner product, $\boldsymbol{e}_k$ is the entity embedding at step $k$, $\boldsymbol{r}_{k''} \in \{\boldsymbol{r}_{kg} \cup \boldsymbol{r}_{interact}  \}$ is the relation embedding at $k''$-th hop, and $\boldsymbol{b}_{{e}_{k+1}}$ is the entity embedding bias.

\textbf{Transition.} Given a state $\Mat{s}_k$ and an action $a_k$, the transition $\delta$ to the next state $\Mat{s}_{k + 1}$ is defined as:
\begin{eqnarray}
\Mat{s}_{k+1} = \delta(\Mat{s}_k, a_k) = \{ u, e_{k-k'},...,  r_{k}, e_k, 
r_{k+1}, e_{k+1} \}.
\end{eqnarray}

\textbf{Reward.} There is no pre-known targeted item for any user in the recommendation, especially in time-aware explainable recommendation.
Instead of setting a binary reward indicating whether the agent has reached a target or not, this paper adopts a soft reward for the terminal state $\Mat{s}_K = (u, h_K, e_K)$ based on the time-aware scoring function $g_R(\hat{v}\mid u)$:
\begin{equation}\label{rewardeq}
R_{K}=\dfrac{g_R (e_{K} \mid u )}{\max _{v \in \mathcal{V}} 
g_R(v \mid u)}
\end{equation}
where $e_{K} \in \mathcal{V}$, and the value of $R_K$ is limited to the range of [0,1]. $g_R(\hat{v} \mid u)$ will be introduced in the following Section \ref{rewardShape}.

\subsubsection{\textbf{Time-aware Based Reward Shaping}}\label{rewardShape}
According to Equation (\ref{rewardeq}), the agent receives a soft reward based on the scoring function $g_R(\hat{v} | u)$. Intuitively, a widely-used  multi-hop scoring function \cite{pgpr} to shape the terminal reward is: 
\begin{equation}
g(\hat{v} \mid u) = (\boldsymbol{e}_u + \boldsymbol{r}_{interact}) \cdot 
\boldsymbol{e}_{\hat{v}} + \boldsymbol{b}_{e_{\hat{v}}},
\end{equation}
where $\cdot$ is the inner product, $u \in \mathcal{U}$ is the target user, $\hat{v} \in \hat{\mathcal{V}}$ is the predicted item at path $\tau_{u, \hat{v}}$ terminal, and $\boldsymbol{b}_{e_{\hat{v}}} \in \mathbb{R}$ is the embedding bias of entity $\hat{v}$. This scoring function is based on an assumption: if there exists a $k$-hop connection between user $u$ and item $\hat{v}$ (say, $\{u\stackrel{r_{1}}{\longrightarrow} e_{1} \stackrel{r_{2}}{\longrightarrow} \ldots \stackrel{r_{k-1}}{\longrightarrow} e_{k-1} \stackrel{r_{k}}{\longrightarrow} \hat{v}\}$), it is reasonable to infer a potential interaction relation $r_{interact}$ between $u$ and $\hat{v}$, which is $u\stackrel{r_{interact}}{\longrightarrow} \hat{v}$.

Nevertheless, this scoring function $g(u, \hat{v})$ hardly works in the TCKG recommendation scenario, because we cannot directly determine which time-aware interaction relation $ r_{interact} \in \mathcal{V}_{\mathcal{U}}^T = \{  \mathcal{V}_{\mathcal{U}}^1, \mathcal{V}_{\mathcal{U}}^2, \cdots, \mathcal{V}_{\mathcal{U}}^{L} \}$ to use for a user $u$.
To address this issue, we design a personalized interaction relation $r_{\hat{v}_u^T}$ for each user $u$ based on her history $h_u$:
\begin{equation}\label{perre}
    r_{\hat{v}_u^T} \leftarrow \boldsymbol{W}_{h_u}\mathcal{V}_\mathcal{U}^T \ ,
\end{equation}
where $\boldsymbol{W}_{h_u} = [w_{h_u}^1, w_{h_u}^2, \cdots, w_{h_u}^L]^{\mathsf{T}}$ is the time cluster weight derived from user $u$'s interaction history $h_u = \{v_u^1, v_u^2, \cdots, v_u^q\}$, where $q$ is the length of $h_u$. 
Here we adopt a statistical method to weight each cluster, we leave the exploration of other weighting methods (\eg, attention mechanism) in future work.
The $l$-th interaction relation's weight $w_{h_u}^l$, in which $l \in \{1,2,...,L\}$, is formulated as follows:
\begin{equation}
w_{h_u}^l = \dfrac{\sum_{i=1}^q{\mathbb{I}(v_u^i = \mathcal{V}_\mathcal{U}^l)}}{q} \ ,
\end{equation}
where $\mathbb{I}(\cdot)$ is the indicator function. A larger relation's weight $w_{h_u}^l$ indicates that interaction $\hat{v}_u^l$ appears more frequently in history $h_u$.
Finally, the personalized time-aware reward
scoring function can be obtained by:
\begin{equation}\label{reward}
g_R(\hat{v} \mid u) = (\boldsymbol{e}_u + \boldsymbol{r}_{\hat{v}_u^T}) \cdot \boldsymbol{e}_{\hat{v}} + \boldsymbol{b}_{\hat{v}} \ .
\end{equation}

\subsubsection{\textbf{Optimization}}
The path reasoning policy is parameterized by using the state information and action space. As illustrated in Section \ref{secMDP}, the component state $\Mat{s}_k = (u, h_k, e_k)$ at step $k$ contains search history $h_k$. Since the reasoning hops may imply dependencies (\eg, temporal dependency or sequence dependency), we introduce a bidirectional LSTM to encode the state vector $\Mat{s}_t$:
\begin{align}
 \Mat{x}_k = & \ \text{dropout}(\sigma(BLSTM(\boldsymbol{e}_u, \boldsymbol{e}_{h_k}, \boldsymbol{e}_k) \ \boldsymbol{W}_1))
\end{align}
where the path reasoning starts from $\Mat{s}_0 = (u, u, \varnothing)$, and for those path lengths with less than $k$-hop historical interactions, we fill zero paddings in their representations;
$\boldsymbol{W}_1$ is linear parameters.
The policy network $\pi$ is an actor and outputs the probability of each action in the pruning action space $\mathcal{A}_k(u)$, and the value network $\hat{c}$ maps the state vector $x_t$ to a real value as the baseline in RL. 
\begin{equation}
\begin{split}
	\pi\left(\cdot \mid \mathbf{s}_k, \mathcal{A}_{k}(u)\right) &=\text{softmax} \left(\mathcal{A}_{k}(u) \times \left(\mathbf{x}_k \boldsymbol{W}_{a}\right)\right) \\
	\hat{c}(\mathbf{s}_k) &=\mathbf{x}_k \boldsymbol{W}_{c}
	\end{split}
\end{equation}
where $\boldsymbol{W}_{a}, \boldsymbol{W}_{c}$ are linear parameters to learn.
These two networks are trained by maximizing the expected reward for any user $u$ in TCKG $\mathcal{G}_K$:
\begin{equation}
	J(\theta) = \mathbb{E}_{a_1, ...,a_K \thicksim 
	\pi}\lbrack\sum_{t=0}^{K-1}\gamma^t 
	R_{k+1} \mid \Mat{s}_0=(u,u,\varnothing)\rbrack
\end{equation}
The optimization is performed via the REINFORCE \cite{reinforcement} algorithm, which updates model parameters $\Theta$ with the following policy gradient: 
\begin{equation}
	\nabla_{\Theta} J(\Theta)=\mathbb{E}_{a_1, ...,a_K \thicksim 
		\pi}\left[\nabla_{\Theta} \log \pi_{\Theta}\left(\cdot \mid \Mat{s}_k, 
		\mathcal{A}_{k}(u)\right)(G-\hat{c}(\Mat{s}_k))\right]
\end{equation}
where $G$ represents the discounted cumulative reward from state $\Mat{s}_k$ to the terminal state $\Mat{s}_K$.

\subsection{Model Prediction}
We now introduce the procedure of temporal ranking recommendation, answering the question ``how to make appropriate time-aware recommendations''.
Similar to the TCKG relation extractor module (Section \ref{tire}), the temporal ranking recommendation module uses the temporal feature space $\mathcal{T} \in \mathbb{R}^m$ to get the time signal.
As illustrated in Figure \ref{pipeline}, the recommend time $\hat{T}$ will be mapped into the temporal feature space $\mathcal{T}$, and we can get the projected recommend temporal vector $\boldsymbol{\hat{T}} \in \mathbb{R}^m$. Then the GMM module will return the clusters distribution $\boldsymbol{W}_{\hat{T}} \in \mathbb{R}^L$ according to $\boldsymbol{\hat{T}}$. Thereafter, the recommendation relation can be calculated by the following formulation:
\begin{equation}\label{rankeq}
\boldsymbol{r}_{W_{\hat{T}}} \leftarrow \boldsymbol{W}_{\hat{T}}\mathcal{V}_\mathcal{U}^T \ ,
\end{equation}

With the input of $u$, the trained time-aware path reasoning module (Section \ref{rewardShape}) will predict the recommended item set $\hat{\mathcal{V}}$ and the corresponding reasoning paths $\tau_{u, \hat{v}}$. Ultimately, we  \textit{rank} the item set $\hat{\mathcal{V}}$ and their paths $\tau_{u, \hat{v}}$ to determine a temporal order to be presented:
\begin{equation}
\text{Rank}(\hat{\mathcal{V}}) = \text{Sort}((\boldsymbol{e}_u + \boldsymbol{r}_{W_{\hat{T}}}) \cdot \boldsymbol{e}_{\hat{\mathcal{V}}})
\end{equation}
where the function $\text{Sort}(\cdot)$ sorts the inner product results in descending order.

%% file: 4-Experiment.tex
\section{Experiments}
\label{exper}

In this section, we conduct experiments to evaluate the performance of our proposed \tkgrec.  Our experiments are intended to address the following research questions:

\begin{itemize}
\item \textbf{RQ1:} How does the proposed \tkgrec~ perform as compared with state-of-the-art normal test and sequence-based recommendation methods?

\item \textbf{RQ2:} How do different components in \tkgrec~ (\textit{\ie,
temporal features,
personalized time-aware reward and
bidirectional LSTM encoder}) affect \tkgrec's performance?

\item \textbf{RQ3:} How do the core parameters ( \ie, \textit{number of time clusters}, \textit{action space size} and \textit{
state history length}) in the time-aware path reasoning module affect the recommendation performance?

\item \textbf{RQ4:} Does the temporal information indeed boost the quality of explainable recommendation?
\end{itemize}

\begin{table}[t!]
\centering
\small
\caption{The statistics of datasets.}
\scalebox{0.78}{
\begin{tabular}{c|c|c|c||c|c|cc}
\hline \hline
\multirow{2}{*}{\textbf{Datasets}} & \multirow{2}{*}{\textbf{Clothing}} & \multirow{2}{*}{\textbf{Cell Phones}} & \multirow{2}{*}{\textbf{Beauty}} & \multirow{2}{*}{\textbf{Relation Details}} & \multirow{2}{*}{\textbf{Description}} & \multicolumn{2}{c}{\textbf{Relation Status}} \\ \cline{7-8} 
                          &                           &                              &                         &                            &                              &  Time-aware?         &  Static?       \\ \hline
\#Users                   & 39,387                    & 27,879                       & 22,363                  &         \#$Purchase$                   &     $User\stackrel{Purchase}{\longrightarrow}Item$ 
&   \checkmark              &   $\times$                \\
\#Items                   & 23,033                    & 10,429                       & 12,101       &  \#$Mention$          &      $User\stackrel{Mention}{\longrightarrow}Feature$                       &    \checkmark                          &      $\times$                             \\
\#Interactions            & 278,677                   & 194,439                      & 198,502                 &   \#$Described\_by$ &      $Item\stackrel{Described\_by}{\longrightarrow}Feature$                         &    \checkmark             &     $\times$              \\
\#Time Clusters           & 24                        & 14                           & 14            &          \#$Bought\_together$ &      $Item_i \stackrel{Bought\_together}{\longrightarrow}Item_j$                        &    $\times$            &      \checkmark              \\
\#Relations Types         & 77                        & 47                           & 47                      &             \#$Also\_viewed$ &      $Item_i \stackrel{Also\_viewed}{\longrightarrow}Item_j$                         &    $\times$            &      \checkmark              \\
\#Entity Types            & 5                         & 5                            & 5                       &            \#$Also\_bought$ &      $Item\_i\stackrel{Also\_bought}{\longrightarrow}Item_j$                         &    $\times$            &      \checkmark             \\
\#Entities                & 425,528                   & 163,249                      & 224,074       
  &            \#$Belong\_to$ &      $Item\stackrel{Belong\_to}{\longrightarrow}Category$                         &    $\times$            &      \checkmark             \\  
\#Triplets                & 10,671,090                & 6,299,494                    & 7,832,720         
      &            \#$Produced\_by$ &      $Item\stackrel{Produced\_by}{\longrightarrow}Brand$                         &    $\times$            &      \checkmark              \\
 \hline \hline
\end{tabular}}
            \label{dataset}
\end{table}

\subsection{Experimental Settings}


\subsubsection{Datasets.}
\label{testdata}
Our experiments are conducted on real-world e-commerce datasets from Amazon ~\cite{datasetAmazon,
HM+2016}, where each dataset contains user interactions (\ie, reviews) and product metadata (\eg, descriptions, brand, price, features, etc) on one specific product category. Here we adopt three typical product categories (\ie, \textit{Clothing}, \textit{Cell Phones} and \textit{Beauty}) and take the standard 5-core version for experiments. For normal test comparison, we follow the previous work \cite{pgpr, jrl} to construct the knowledge graph. Also, we use the same training and test sets as that of \cite{pgpr}. That is, we randomly sample 70\% of user purchases for model training and take the rest 30\% for testing. From the training set, we randomly select 10\% of interactions as validation set to tune hyperparameters. The statistics of each dataset are presented in Table \ref{dataset}, we have three kinds of \textit{Time-aware} relations and five kinds of \textit{Static} relations. The total number of \textit{Relations Types} is obtained by calculating the number of \textit{Time-aware} relations times the number of \textit{Time Clusters} and plus the number of \textit{Static} relations.
We also utilize timestamps of reviews for temporal study. Specifically, we extract time-aware interaction relations as described Section~\ref{tire}.
For sequence-based comparison\cite{seqd1, seqd2, sasrec}, we take the first 60\% of interactions in each user's purchase sequence, and use the next 10\% of interactions as the validation set to tune hyperparameters for all sequential recommendation methods. The rest 30\% of interactions are used as the test set.

\subsubsection{Compared methods.}
We compare the performance of \tkgrec~ with the following baselines, which include KG-based and sequence-based recommendation methods:
\begin{itemize}
\item \textbf{BPR} \cite{bpr}: the classic recommendation method that learns user and item embedding with a pair-wise objective function.
\item \textbf{RippleNet} \cite{ripplenet}: a KG-based recommendation method with modeling preference propagation along with the knowledge graph.
\item \textbf{TransRec} \cite{transRec}: a translation-based recommendation method that map both user and item representations in a specific shared space with
        personalized translation vectors.
%
%
\item \textbf{PGPR} \cite{pgpr}: a KG-based explainable recommendation method that performs path reasoning on knowledge graph to make recommendation as well as provide interpretations.
\item \textbf{ADAC} \cite{ADAC}: a state-of-the-art explainable recommendation method that extends PGPR with leveraging demonstrations to guide path finding.

\item \textbf{GRU4Rec} \cite{gru4rec}: a session-based recommendation method equipped with RNN structured GRU.

\item \textbf{BERT4Rec} \cite{bert4rec}: a bidirectional self-attention framework that models user behaviors.

\item \textbf{GC-SAN} \cite{gcsan}: a graph contextualized self-attention model that utilizes GNN and self-attention for session-based recommendation.

\item \textbf{NextItNet} \cite{nextit}: a session-based CNN recommender that captures short- and long-range item dependencies.

\item \textbf{SASRec} \cite{sasrec}: a self-attention based sequential model that utilizes Markov Chains and RNNs to capture short-term and long-term user behaviors.

\item \textbf{SR-GNN} \cite{srgnn}: a session-based recommendation method equipped with Graph Neural Networks.

\item \textbf{TimelyRec} \cite{heteTemporal}: a timely recommendation method that jointly considers periodic and evolving heterogeneous temporal patterns of user preferences, and proposes a cascade of two encoders to capture such patterns.

\item \textbf{\tkgrec}: the method proposed in this work. We mainly test three versions of \tkgrec:  w/ $\boldsymbol{f_{stru}}$ that removes statistical features and only uses structural features in the time-aware relation extraction module; w/ $\boldsymbol{f_{stat}}$ that only uses statistical features; \tkgrec\ that uses both statistical and structural features.
%

\end{itemize}

\begin{figure}[!t]
    \begin{minipage}[b]{\columnwidth}
        \centering
        \subfloat[BIC value with the clustering number.]{
            \centering
            \includegraphics[width=0.43\columnwidth]{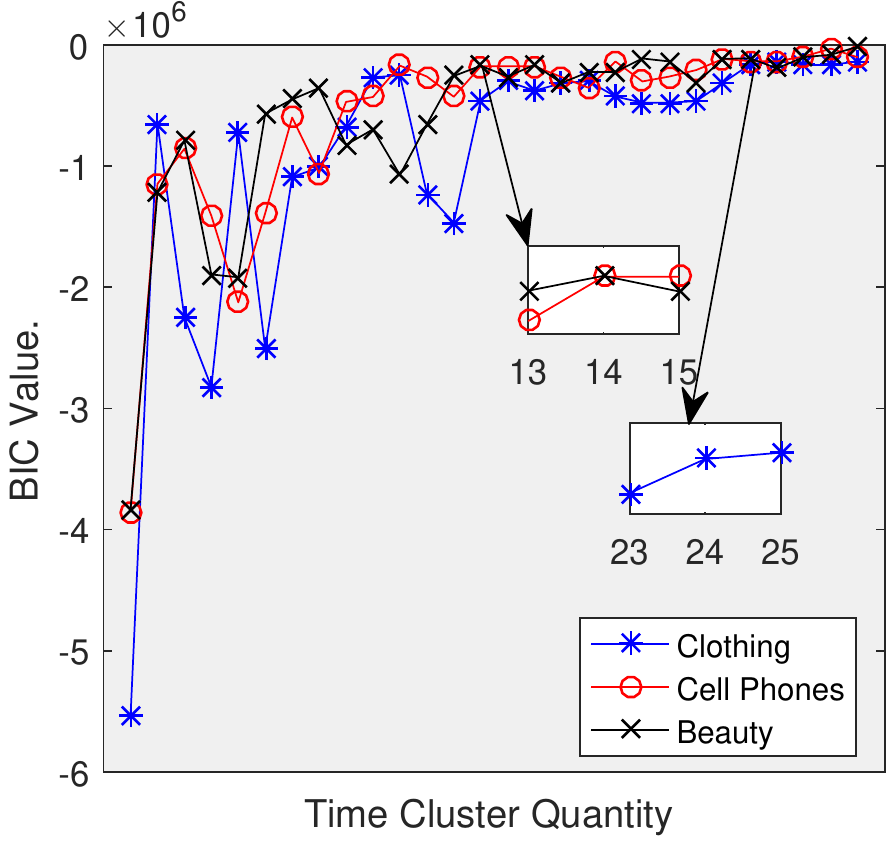}
            \label{bic}
        }
        \subfloat[The number of invalid users.]{
            \centering
            \includegraphics[width=0.5\columnwidth]{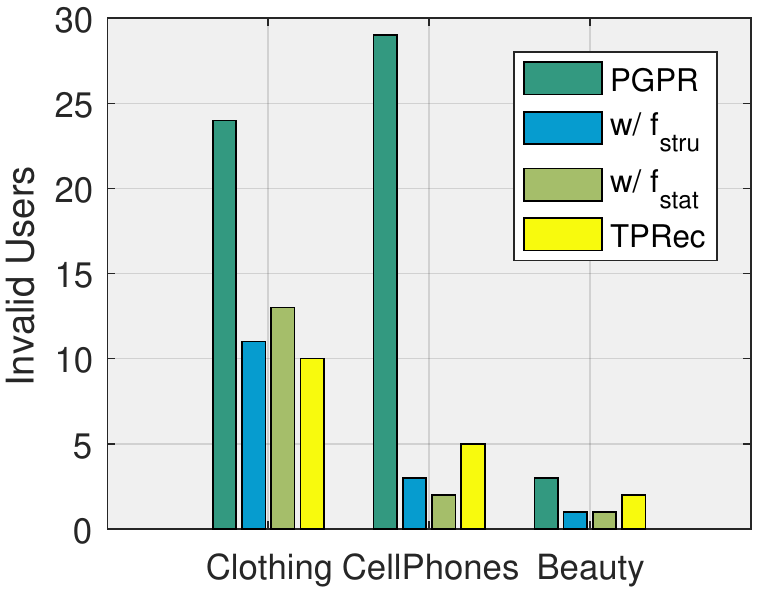}
            \label{invalid}
        }
        \caption{Analyses of different datasets. The left subgraph illustrates how BIC value varies with the clustering number, while the right subgraph illustrates the numbers of invalid users calculated from PGPR and \tkgrec.}
    \end{minipage}
    \vspace{-0.15cm}
\end{figure}

\subsubsection{Implementation Details.}\label{parasetting}
%

To reduce the experiment workload and keep the comparison fair, we closely follow the settings of the PGPR work ~\cite{pgpr}.
For the time-aware relation extraction module, we perform relation clustering with GMM and select the optimal number of clusters with Bayesian Information Criterion (BIC) ~\cite{bic}. The BIC value with the clustering number is presented in Figure \ref{bic}. The optimal clustering numbers are 24, 14, and 14 on \textit{Clothing}, \textit{Cell Phones} and \textit{Beauty}, respectively.

For the time-aware representation learning module, we set the embedding size $d$ as $100$, and the dimension $m$ of temporal feature space as $25$, with 16 dimensions for statistical features and 9 dimensions for structural features. More specifically, for statistical features, we 
simply compute the $year(t) \in \mathbb{R}^1$ to be $= current\ year - earliest\ year$, $day(t) \in \mathbb{R}^1 = current\ day - earliest\ day$, $month(t) \in \mathbb{R}^1$ is the month of the year, $week(t) \in \mathbb{R}^8$ is the day in a week and it's one-hot encoding, and $season(t) \in \mathbb{R}^5$ is the season in a year and it's one-hot encoding. For structural features, given a specific timestamp $t_1$, we have $\boldsymbol{f}_{stru}(t_1) \in \mathbb{R}^9$ $= z(t_1)\ ||\ z_{sea}'(t_1)\ ||\ z_{sea}''(t_1)$ $||\ 
\ z_{mon}'(t_1)\ ||\ z_{mon}''(t_1)$ $||\ z_{week}'(t_1)\ ||\ z_{week}''(t_1)
$ $||\ z_{day}'(t_1)\ ||\ z_{day}''(t_1)$.

For the time-aware path reasoning module, we set the maximum episode path length $k$ as $3$ (namely,
the maximum history length is $2$), the default history length $k'$ as 1, and the maximum size of pruned action space as $|\mathcal{A}_k|\leq250$.
Besides, for the parameters in RL, we set the dimension of $x_t$ as 256 and the discount factor $\gamma$ as $0.99$; Also, for the policy/value network, we set $W_1 \in \mathbb{R}^{512 \times
256}$ , $W_a \in \mathbb{R}^{256 \times 251}$ and $W_c \in \mathbb{R}^{256 \times 1}$. We optimize our model for 50 epochs with Adam, where the learning rate is set as 0.0001 and the batch size is 32. During the prediction, we refer to PGPR \cite{pgpr} and set the sampling sizes as $K_1=25$, $K_2=5$, $K_3=1$.

%

%

\subsubsection{Evaluation Metrics.}\label{evalMetric}
We adopt four widely used metrics to evaluate recommendation performance including: 
\textit{Normalized Discounted Cumulative Gain@K} (\textit{NDCG@K} for short), 
\textit{Recall@K}, \textit{Precision@K} and \textit{Hit Ratio@K} (\textit{HR@K} for short).
The process for obtaining the scores for those metrics is given in Equation (\ref{metric_fun}).

\begin{align}   \label{metric_fun}
	 NDCG@K &= \frac{DCG@K}{IDCG@K} & Recall@K &= \frac{\left| TP \right|}{\left| TP \right| + \left| FN \right|} \notag \\
	 Precision@K &= \frac{\left| TP \right|}{\left| TP \right| + \left| FP \right|} & HR@K &=  \frac{N_{hits} @K}{N_{users}}
\end{align}
Here, \textit{DCG@K} is the  Discounted Cumulative Gain at position K, \textit{IDCG@K} is Ideal Discounted Cumulative Gain through position K, $\left| TP \right|$ means the number of recommended items that are actually relevant at position K, $\left| TP \right| + \left| FN \right|$ is the total number of relevant items and  $\left| TP \right| + \left| FP \right|$ means the number of recommended items at position K. As for \textit{HR@K}, $N_{hits} @K$ is the number of users whose items in the test set at position K, and $N_{users}$ is the total number of users.

These metrics reflect how a model retrieves relevant items that users are indeed fond of. A larger score indicates the better performance that a model achieves. In our experiments, we refer to ~\cite{pgpr} and set $\text{K}=10$.

\begin{table}[t!]
\centering
\small
\caption{Normal test performance comparison on three real-world datasets. The results are reported in percentage (\%). The bold font denotes the winner in that column. The row `Impv' indicates the relative performance gain of our \tkgrec~ relative to ADAC.}\label{overallComp}
{
\begin{tabular}{c|p{0.65cm}<{\raggedleft}p{0.65cm}<{\raggedleft}p{0.65cm}<{\raggedleft}p{0.65cm}<{\raggedleft}|
p{0.65cm}<{\raggedleft}p{0.65cm}<{\raggedleft}p{0.65cm}<{\raggedleft}p{0.65cm}<{\raggedleft}|
p{0.65cm}<{\raggedleft}p{0.65cm}<{\raggedleft}p{0.65cm}<{\raggedleft}p{0.65cm}<{\raggedleft}}
\hline  \hline
Models      &
\multicolumn{4}{c|}{\textbf{Clothing}}                               &
\multicolumn{4}{c|}{\textbf{Cell Phones}}                                 &
\multicolumn{4}{c}{\textbf{Beauty}}
\\ \hline
Metrics     & \textbf{NDCG}  & \textbf{Recall} & \textbf{HR}     &
\textbf{Prec.} & \textbf{NDCG}  & \textbf{Recall} & \textbf{HR}    &
\textbf{Prec.} & \textbf{NDCG} & \textbf{Recall} & \textbf{HR}     &
\textbf{Prec.} \\ \hline
BPR         & 0.598          & 1.086           & 1.801          &
0.196              & 1.892          & 3.363           & 5.323           &
0.624              & 2.704         & 4.927           & 9.113           &
1.066              \\
RippleNet   & 0.627          & 1.112           & 1.885          &
0.201              & 1.935          & 3.858           & 5.727           &
0.688              & 2.458         & 5.251           & 9.224           &
1.133              \\
TransRec    & 1.245          & 2.078           & 3.116          &
0.312              & 3.361          & 6.279           & 8.725           &
0.962              & 3.218         & 4.853           & 8.671           &
1.285              \\
PGPR        & 2.871          & 4.827           & 7.023          &
0.723              & 5.042          & 8.416           & 11.904          &
1.274              & 5.573         & 8.476           & 14.682          &
1.744              \\
ADAC        & \underline{3.048}          & \underline{5.027}           & \underline{7.502}          &
\underline{0.763}              & \underline{5.220}           & \underline{8.943}           & \underline{12.537}          &
\underline{1.358}              & \underline{6.080} & \underline{9.424}  & \underline{16.036} &
\underline{1.991}     \\  \hline
w/ $\boldsymbol{f_{stru}}$        & 3.074          & 5.132           &
7.484          &
0.778              & \textbf{5.988}          & \textbf{10.185}           &
\textbf{14.278}          & \textbf{1.528}              & \textbf{6.185}
& \textbf{9.582}     & \textbf{16.130}          &
\textbf{2.023}              \\
w/ $\boldsymbol{f_{stat}}$        & 2.997          & 5.144           &
7.498          &
0.781              & 5.743          & 9.714           & 13.649          &
1.471              & 5.868         & 9.227           & 15.719          &
1.928              \\
\tkgrec\ & \textbf{3.109} & \textbf{5.298}  & \textbf{7.657} &
\textbf{0.798}    &5.643 & 9.572 & 13.553
&1.463     & 5.963 & 9.184 & 15.678 & 1.919              \\   \hline
Impv.(\%)    & +2.01          & +5.39           & +2.07          &
+4.59             & +14.71         & +13.89         & +13.89          &
+12.52            & +1.73 & +1.68 & +0.59  & +1.66            \\  \hline  \hline
\end{tabular}
}
\end{table}

\begin{table}[t!]
\caption{Sequence-based performance comparison. The results are reported in percentage (\%). }\label{seqComp}
\begin{tabular}{c|cc|rr|rr|rr}
\hline \hline
\multirow{2}{*}{\textbf{Models}} & \multicolumn{2}{c|}{\textbf{Side Infomation}} & \multicolumn{2}{c|}{\textbf{Clothing}}                                   & \multicolumn{2}{c|}{\textbf{Cell Phones}}                                & \multicolumn{2}{c}{\textbf{Beauty}}                                     \\ \cline{2-9} 
                         & \multicolumn{1}{c}{\textbf{KG?}}       & \multicolumn{1}{c|}{\textbf{Time?}}      & \multicolumn{1}{c}{\textbf{NDCG}} & \multicolumn{1}{c|}{\textbf{Recall}} & \multicolumn{1}{c}{\textbf{NDCG}} & \multicolumn{1}{c|}{\textbf{Recall}} & \multicolumn{1}{c}{\textbf{NDCG}} & \multicolumn{1}{c}{\textbf{Recall}} \\ \hline
BERT4Rec  &   $\times$                 &          \checkmark                & 0.248                             & 0.516                                & 1.612                             & 3.315                                & 1.664                             & 3.540                               \\
GRU4Rec  &   $\times$                 &          \checkmark                         & 0.541                             & 1.070                                & 3.370                             & 6.365                                & 3.813                             & 6.508                               \\
NextItNet &   $\times$                 &          \checkmark                         & 0.289                             & 0.573                                & 2.422                             & 4.683                                & 2.360                             & 4.727                               \\
SR-GNN   &   $\times$                 &          \checkmark                          & 0.428                             & 0.797                                & 3.072                             & 5.777                                & 3.212                             & 5.757                               \\
GC-SAN    &   $\times$                 &          \checkmark                         & 1.116                             & 2.229                                & \underline{4.211}                       & \underline{8.219}                          & 4.361                             & 7.511                               \\
SASRec  &   $\times$                 &          \checkmark                         & 1.253                             & 2.541                                & 3.876                             & 7.695                                & \underline{4.522}                       & \underline{8.918}                         \\
TimelyRec &   $\times$                 &          \checkmark                         & 1.263                             & 2.716                                & 4.039                             & 7.791                                & \textbf{4.671}                    & \textbf{9.217}                      \\
PGPR      &     \checkmark               &      $\times$                    & \underline{2.102}                       & \underline{3.651}                          & 3.135                             & 4.911                                & 3.654                             & 6.060                               \\ \hline
w/ $\boldsymbol{f_{stru}}$      &     \checkmark               &       \checkmark                   & 2.245                             & 3.905                                & 4.228                             & 8.936                                & 4.331                             & 8.051                               \\
w/ $\boldsymbol{f_{stat}}$       &     \checkmark               &       \checkmark                        & 2.270                             & 3.955                                & 4.123                             & 8.767                                & 4.175                             & 7.717                               \\
\tkgrec     &     \checkmark               &       \checkmark                         & \textbf{2.304}                    & \textbf{4.027}                       & \textbf{4.292}                    & \textbf{9.131}                       & 4.467                             & 8.773                               \\ \hline \hline
\end{tabular}
\end{table}

\subsection{Performance Comparison (RQ1)}
\label{rq1}
Table \ref{overallComp} presents the performance of the compared normal test methods in terms of four evaluation metrics.  For the sake of
clarity, the row `Impv' shows the relative improvement achieved by our \tkgrec\ or w/ $\boldsymbol{f_{stru}}$ over the best baselines. 
Table \ref{seqComp} presents the sequence-based recommendation methods performance comparison.
The boldface font denotes the winner in that column.
We make the following observations:

\subsubsection{\textbf{\tkgrec~ Vs. existing normal test methods.}}
From Table \ref{overallComp}, we observe our \tkgrec~ always achieve the best performance among all compared methods. Especially in the dataset Cell Phones, the improvement of w/ $\boldsymbol{f_{stru}}$ over best baselines are quite impressive --- 14.71\%, 13.89\%, 13.89\%, and 12.52\% in terms of NDCG, Recall, HR, and Precision, respectively. 
%
%
And our \tkgrec~ performs significantly better ($p$-value \textless\ 0.05) than those open source baselines (PGPR, TransRec, \textit{etc}.) with three repeated trials.
This result validates that by leveraging temporal information our \tkgrec~ is able to find higher-quality reasoning paths.  

We make a further investigation to provide more insights. We refer to a path as an \textit{invalid path} if the path starts from a user but doesn't end at an item entity within three hops (note that we set the path length $k=3$). We also refer to a target user that we predict  as an \textit{invalid user} if he has less than 10 \textit{valid} paths. As shown in Figure \ref{invalid}, we find the number of \textit{Invalid Users} in \tkgrec\ is consistently less than PGPR which does not consider temporal information. This result suggests that time-aware rewards adopted by \tkgrec\ indeed guide the RL agent towards exploration better.

\subsubsection{\textbf{Comparison in terms of different datasets.}}
 Interestingly, in normal test setting, we observe that the improvement of our \tkgrec\ over recent work on the datasets Clothing and Cell Phones are relatively large, while relatively small on the dataset Beauty. This result can be explained by the difficulty of different datasets. From Figure \ref{invalid}, we observe that the number of invalid users on the Beauty dataset generated by PGPR is $3$, while the number increases to $24$ and $29$ on other datasets. This result validates that path reasoning on Beauty is relatively easier than on other datasets. PGPR or ADAC also achieve pretty good performance on Beauty. Correspondingly, our \tkgrec\ does not obtain much gain on Beauty.

\subsubsection{\textbf{Reasoning Vs. Unreasoning.}}
 We also observe that KG-based reasoning methods (\ie, PGPR, ADAC and \tkgrec) outperform other KG-based methods (\ie, RippleNet) on all three datasets across all four evaluation metrics. This result clearly validates the superiority of conducting path reasoning on knowledge graph. Though reasoning, interpretability is not against accuracy. It not only increases interpretability, but also boosts model accuracy.
 
 \subsubsection{\textbf{\tkgrec~ Vs. existing sequence-based  methods.}}
 From Table \ref{seqComp}, we can see the performance of \tkgrec\ and PGPR is lower than that at normal test setting.
One possible reason of the performance drop is the Out-of-Distribution (OOD) setting in the sequence-based experiment, which means the test distribution is different from the training.
%
Our \tkgrec\ achieves the best performance on the \textit{Clothing} and the \textit{Cell Phones} datasets, whereas, slightly worse than TimelyRec and SASRec on the \textit{Beauty} dataset.
It is appropriate to emphasize that these sequential recommendation methods are specifically designed for predicting the next item that the user would interact, which sacrifices explainability. Our \tkgrec\ can not only provide explanations about why an item is suitable for a user, but also achieve comparable recommendation accuracy in sequential recommendation.

We further make the following observations. In the \textit{Clothing} dataset, even PGPR is much better than other sequence-based methods, which indicates that the KG side information is really useful in the \textit{Clothing} dataset. Additionally, our \tkgrec\ consistently beats PGPR in all  datasets, and \tkgrec\ performs better than the other two versions, $w/ f_{stru}$ and $w/ f_{stat}$. 
Notably, there is 10\% of the interactions in validation set between training set and test set in sequence-based comparison, and the time-series clustering method doesn't know the interaction distribution around the test set, which is different from normal test setting. 
Hence, $w/ f_{stru}$'s effect is not enough to support such experiments, it needs to be supplemented by $w/ f_{stat}$.
%
%
This verifies the validity of our time-aware reasoning methodology.

\begin{table}[t!]
\caption{Characteristics and performance of \tkgrec\ and its variants in normal test setting.  The `Imp(\%).' is relative to \tkgrec.}\label{abCmpt}
\resizebox{\textwidth}{27mm}{
\begin{tabular}{c|cc|cc|llllll}
\hline \hline
\multirow{3}{*}{Methods} & \multicolumn{2}{c|}{Reasoning With.}                                                    & \multicolumn{2}{c|}{Equipped with.}                                            & \multicolumn{6}{c}{Performance}                                                                                  \\ \cline{2-11} 
                         & Time-aware                                 & BLSTM                                      & Statistical                                & Structural                                 & \multicolumn{2}{c|}{Clothing}            & \multicolumn{2}{c|}{Cell Phones}         & \multicolumn{2}{c}{Beauty} \\ \cline{6-11} 
                         & Reward?                                    & Encoder?                                   & Feature?                                   & Feature?                                   & Prec.    & \multicolumn{1}{l|}{Recall}   & Prec.    & \multicolumn{1}{l|}{Recall}   & Prec.        & Recall      \\ \hline
\textbf{PGPR}            & \multirow{2}{*}{$\times$}    & \multirow{2}{*}{$\times$}    & \multirow{2}{*}{$\times$}    & \multirow{2}{*}{$\times$}    & 0.723    & \multicolumn{1}{l|}{4.827}    & 1.274    & \multicolumn{1}{l|}{8.416}    & 1.744        & 8.476       \\
Imp(\%)                  &                                            &                                            &                                            &                                            & -9.40  & \multicolumn{1}{l|}{-8.89}  & -12.9 & \multicolumn{1}{l|}{-12.1} & -9.12      & -7.71     \\ \hline
\textbf{w/o R}         & \multirow{2}{*}{$\times$}    & \multirow{2}{*}{\checkmark} & \multirow{2}{*}{\checkmark} & \multirow{2}{*}{\checkmark} & 0.191    & \multicolumn{1}{l|}{1.374}    & 0.317    & \multicolumn{1}{l|}{1.909}    & 0.712        & 3.031       \\
Imp(\%)                  &                                            &                                            &                                            &                                            & -76.1 & \multicolumn{1}{l|}{-74.1} & -78.3 & \multicolumn{1}{l|}{-80.1} & -62.9     & -67.0    \\ \hline
\textbf{w/o L}         & \multirow{2}{*}{\checkmark} & \multirow{2}{*}{$\times$}    & \multirow{2}{*}{\checkmark} & \multirow{2}{*}{\checkmark} & 0.786    & \multicolumn{1}{l|}{5.205}    & 1.450    & \multicolumn{1}{l|}{9.534}    & 1.927        & 9.200       \\
Imp(\%)                  &                                            &                                            &                                            &                                            & -1.50  & \multicolumn{1}{l|}{-1.76}  & -0.89  & \multicolumn{1}{l|}{-0.40}  & +0.42       & +0.17      \\ \hline
\textbf{w/ $\boldsymbol{f_{stru}}$}     & \multirow{2}{*}{\checkmark} & \multirow{2}{*}{\checkmark} & \multirow{2}{*}{$\times$}    & \multirow{2}{*}{\checkmark} & 0.778    & \multicolumn{1}{l|}{5.132}    & 1.528    & \multicolumn{1}{l|}{10.185}   & 2.023        & 9.582       \\
Imp(\%)                  &                                            &                                            &                                            &                                            & -2.51  & \multicolumn{1}{l|}{-3.13}  & +4.44   & \multicolumn{1}{l|}{+6.40}   & +5.42       & +4.33      \\ \hline
\textbf{w/ $\boldsymbol{f_{stat}}$}     & \multirow{2}{*}{\checkmark} & \multirow{2}{*}{\checkmark} & \multirow{2}{*}{\checkmark} & \multirow{2}{*}{$\times$}    & 0.781    & \multicolumn{1}{l|}{5.144}    & 1.471    & \multicolumn{1}{l|}{9.714}    & 1.928        & 9.227       \\
Imp(\%)                  &                                            &                                            &                                            &                                            & -2.13  & \multicolumn{1}{l|}{-2.91}  & +0.55   & \multicolumn{1}{l|}{+1.48}   & +0.47       & +0.47      \\ \hline
\textbf{\tkgrec}      & \checkmark                  & \checkmark                  & \checkmark                  & \checkmark                  & 0.798    & \multicolumn{1}{l|}{5.298}    & 1.463    & \multicolumn{1}{l|}{9.572}    & 1.919        & 9.184       \\ \hline \hline
\end{tabular}}
\end{table}

\begin{figure}[tb]
\includegraphics[width=\columnwidth]{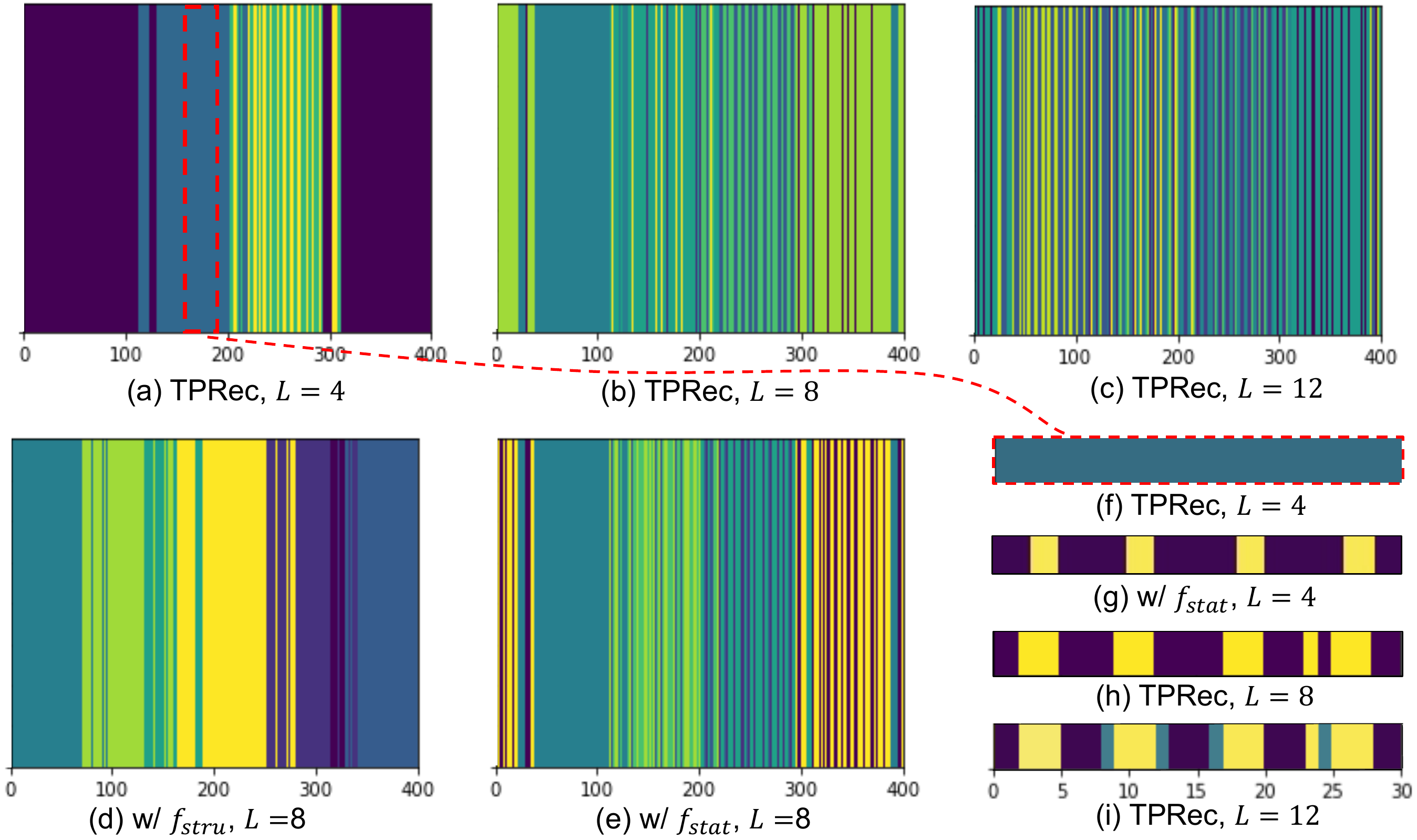}
\caption{ Visualization of the clustering of the relations with varying timestamp on the \textit{Cell Phones} dataset. 
The color represents the class that the timestamp belongs to. 
				}
		\label{cluster}
\end{figure}

\subsection{Ablation Study (RQ2)}

Note that \tkgrec\ improves PGPR mainly in the following three aspects: using temporal features to extract relations, adopting personalized time-aware reward strategy, and using bidirectional LSTM to encode the state vector. To show the impact of these aspects, besides making a comparison among three versions of \tkgrec\ (w/ $\boldsymbol{f_{stru}}$, w/ $\boldsymbol{f_{stat}}$ and \tkgrec ), we further conduct ablation studies and compare \tkgrec\ with its two variants: (1) \textit{w/o R} that removes personalized time-aware weighting strategy $W_{h_u}$ in generating rewards and uses average pooling instead. (2) \textit{w/o L} that replaces bidirectional LSTM with a simple fully connected layer as PGPR.  The characteristics and performance of these methods are presented in Table \ref{abCmpt}.

\subsubsection{\textbf{Effectiveness of using temporal features.}}
 First, we observe our \tkgrec  (\ie, w/ $\boldsymbol{f_{stru}}$, w/ $\boldsymbol{f_{stat}}$ and \tkgrec) that use temporal features consistently outperform PGPR. This result validates the essence of temporal information in making recommendation. We then make a comparison among three versions of \tkgrec. To our surprising, w/ $\boldsymbol{f_{stru}}$ that just uses structure features usually achieves better performance than \tkgrec\ 
which uses both statistical and structural features in normal test setting. To gain more insights of this phenomenon, we visualize the results of the \textit{Cell Phones} dataset  clustering in Figure \ref{cluster}, and the other two datasets have similar performance.
 We observe that with the cluster number $L$ gaining, the clustering result in \tkgrec\ turns to be more complicated. 
 %
 %
 For example, in Figure \ref{cluster}a, the clustering result of \tkgrec\ is similar with the four seasons in one year,
and then we focus on 30 days in the second season (Figure \ref{cluster}f). As the $L$ grew to $8$, it broke the seasons' pattern, and could obtain weekend-like clusters (Figure \ref{cluster}h). It’s worth noting that w/ $\boldsymbol{f_{stat}}$ (Figure \ref{cluster}g) also distinguishes between weekdays and weekends, but our \tkgrec\ clustering is more flexible, which benefits from the structural features. When $L = 12$ in Figures \ref{cluster}c and \ref{cluster}i, the clusters become even more precise. 
 Furthermore, in Figure \ref{cluster}b, \ref{cluster}d and \ref{cluster}e, 
 we can see that the clustering is more stable in w/ $\boldsymbol{f_{stru}}$, where the interactions with similar tendencies are more likely to be clustered into the same class, while we observe relatively heavy fluctuation in w/ $\boldsymbol{f_{stat}}$. 
 %
 %
 This interesting phenomenon suggests that the structural features deduced from user behaviors provide an important and reliable signal in capturing interaction similarity, while statistical features exhibit weak or even noisy correlation with the similarity. Nonetheless, the statistical features sometimes are also useful and serve as a complement for relation extraction. It can be seen from the better performance of \tkgrec\ over w/ $\boldsymbol{f_{stru}}$ on dataset Clothing and all three datasets in sequence-based test setting.

\subsubsection{\textbf{Effectiveness of using time-aware reward.}}
From Table \ref{abCmpt}, when removing personalized time-aware weighting strategy in reward, we observe significant performance degradation (higher than 69.21\%) of w/o R compared with \tkgrec. w/o R even performs worse than PGPR. This interesting phenomenon clearly reveals the importance of using personalized weights. Time-aware relations are indeed useful for path reasoning, but they require to be carefully exploited. Only if we finely differentiate the contribution of each type of relation, can we sufficiently enjoy the merits of such information.


\subsubsection{\textbf{Effectiveness of using bidirectional LSTM}}
By comparing w/o L with \tkgrec, generally speaking, we can conclude that using bidirectional LSTM could boost recommendation performance. This result can be attributed to the powerful capacity of BLSTM in capturing dependency between reasoning hops.  However, on the dataset Beauty, \tkgrec\ performs closely or even worse than w/o L. This phenomenon can be attributed to the specialty of Beauty. As discussed in Section~\ref{rq1}, Beauty is a relatively simple dataset and does not need such complex structure. The advanced BLSTM may not bring much gain on Beauty and instead would increase the risk of over-fitting.

\begin{figure}[]
	\centering
	\subfloat[Clothing.]{
		\begin{minipage}[t]{0.33\textwidth}
			\centering
			\includegraphics[width=1\textwidth]{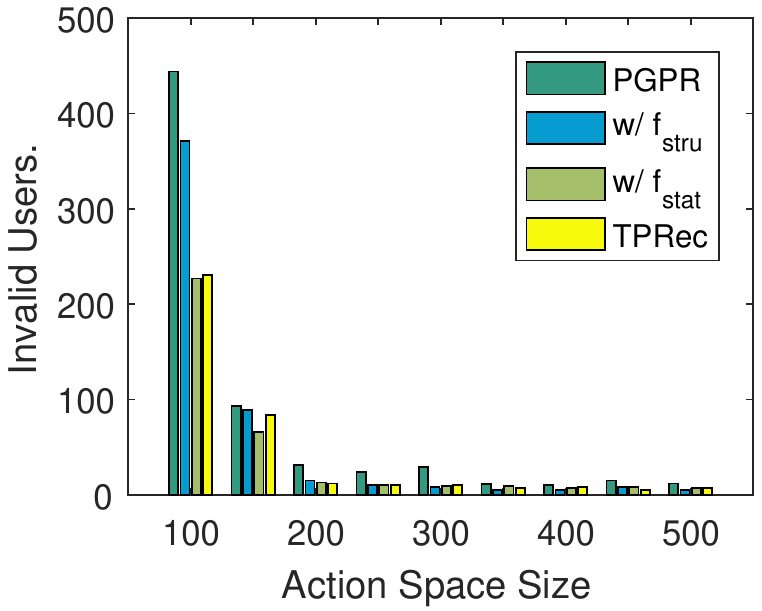}
		\end{minipage}
	}
	\subfloat[Cell Phone.]{
		\begin{minipage}[t]{0.33\textwidth}
			\centering
			\includegraphics[width=1\textwidth]{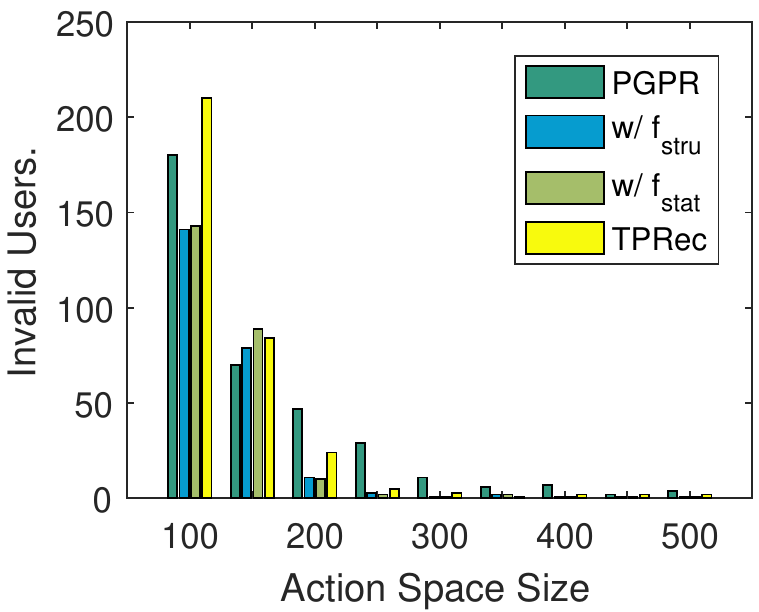}
		\end{minipage}
	}
	\subfloat[Beauty.]{
		\begin{minipage}[t]{0.33\textwidth}
			\centering
			\includegraphics[width=1\textwidth]{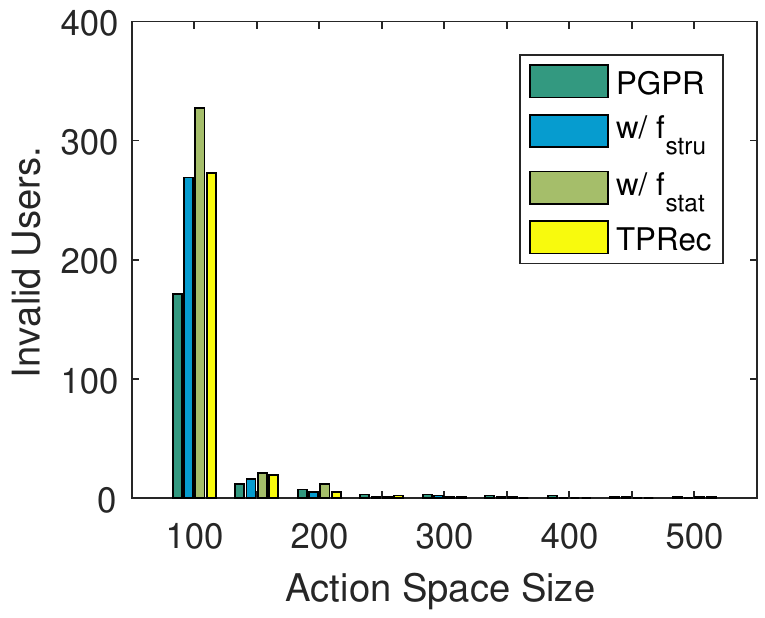}
		\end{minipage}
	}
\caption{ Invalid User number of varying action space size $\epsilon$.
}
	\label{act_invalid}
\end{figure}

\subsection{Parameter Study on Path Reasoning Module. (RQ3)}
To provide more insights on the time-aware path reasoning module, we test the performance of \tkgrec\ with different time cluster quantity, action space size $\epsilon$ and state history length $k'$.

\subsubsection{\textbf{Effect of Time Cluster Quantity.}}
\begin{figure}[]
	\centering
	\subfloat[NDCG (Clothing).]{
		\begin{minipage}[t]{0.25\textwidth}
			\centering
			\includegraphics[width=1\textwidth]{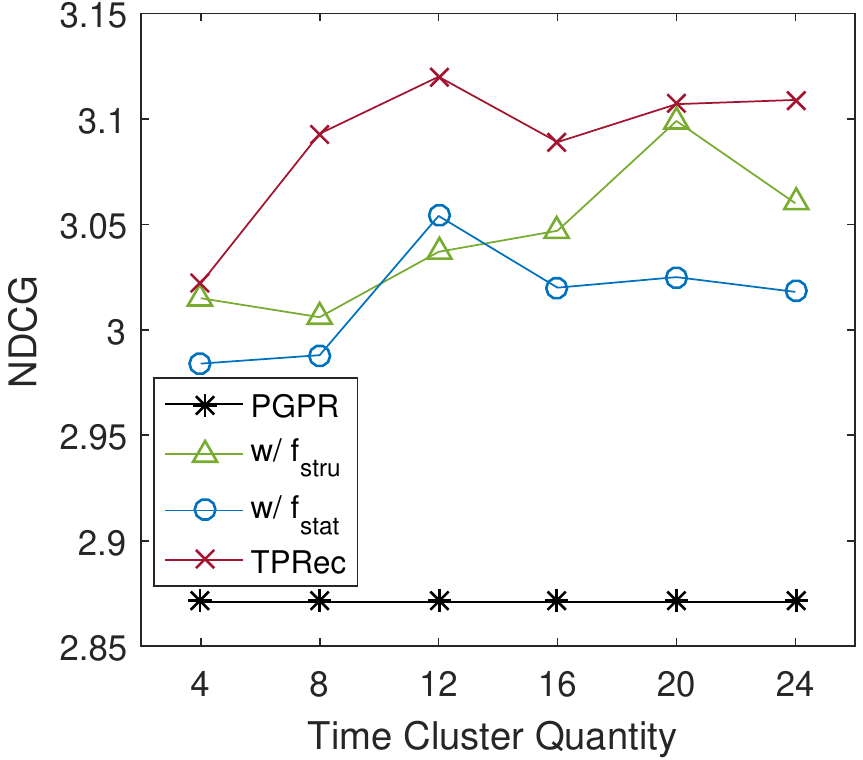}
		\end{minipage}
	}
	\subfloat[Recall (Clothing).]{
		\begin{minipage}[t]{0.25\textwidth}
			\centering
			\includegraphics[width=1\textwidth]{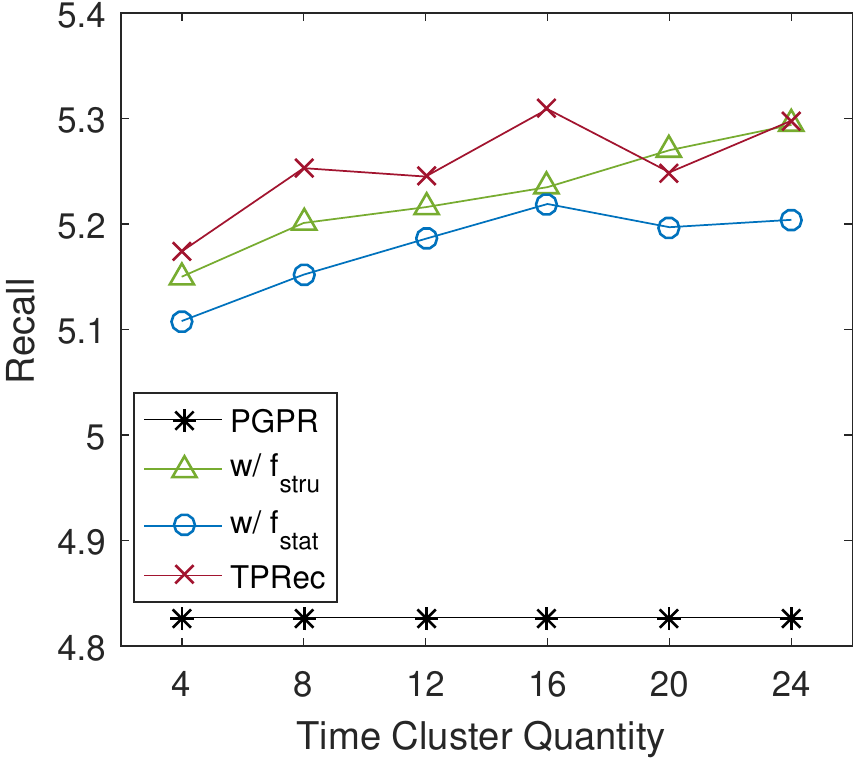}
		\end{minipage}
	}
	\subfloat[HR (Clothing).]{
		\begin{minipage}[t]{0.25\textwidth}
			\centering
			\includegraphics[width=1\textwidth]{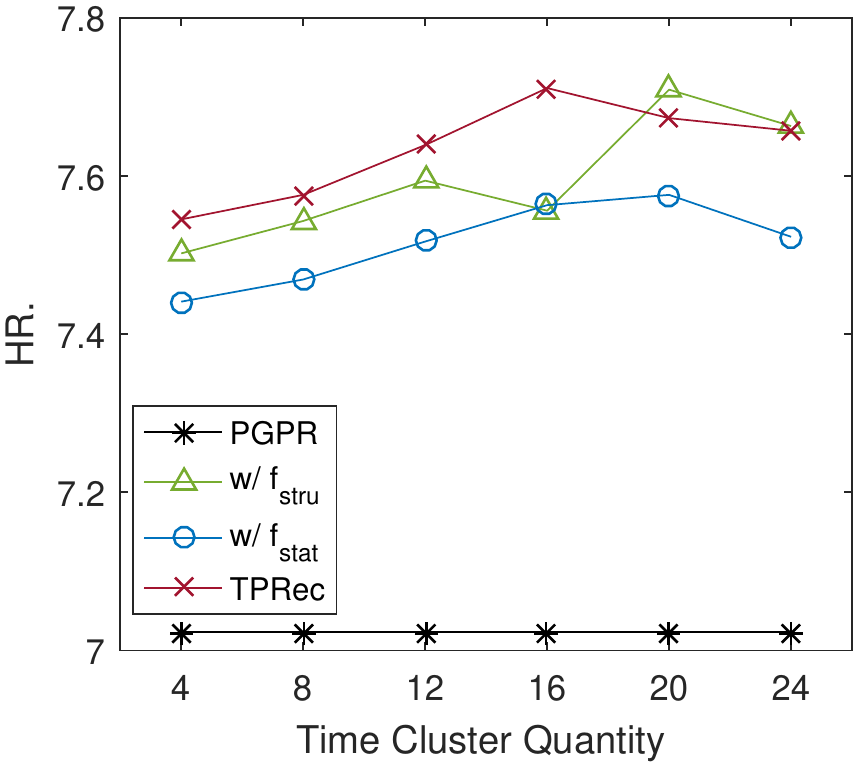}
		\end{minipage}
	}
	\subfloat[Precision (Clothing).]{
		\begin{minipage}[t]{0.25\textwidth}
			\centering
			\includegraphics[width=1\textwidth]{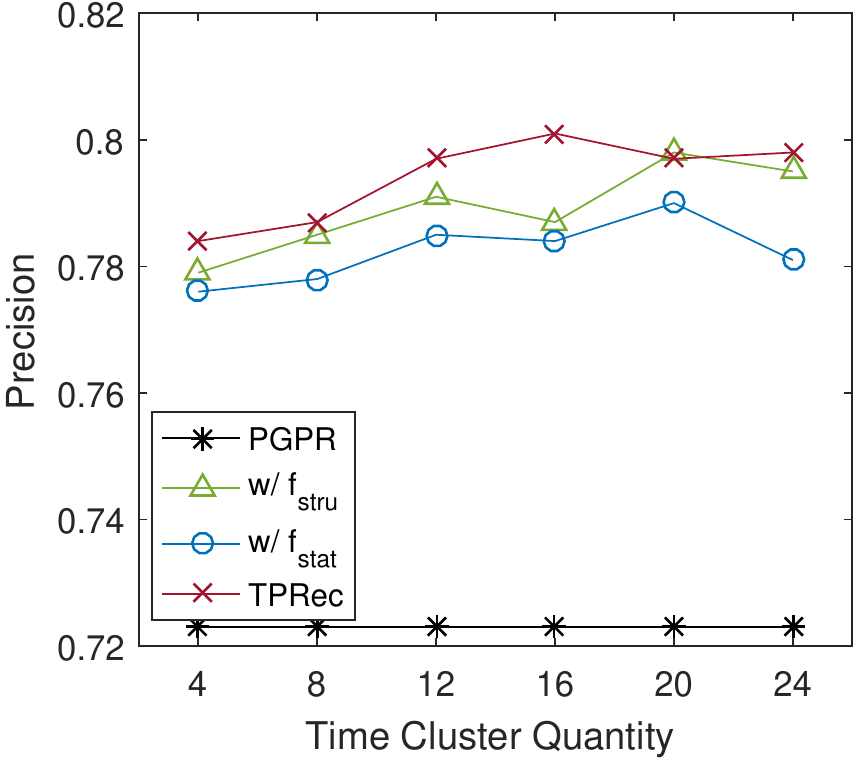}
		\end{minipage}
	}
	
	\subfloat[NDCG (Clothing).]{
		\begin{minipage}[t]{0.25\textwidth}
			\centering
			\includegraphics[width=1\textwidth]{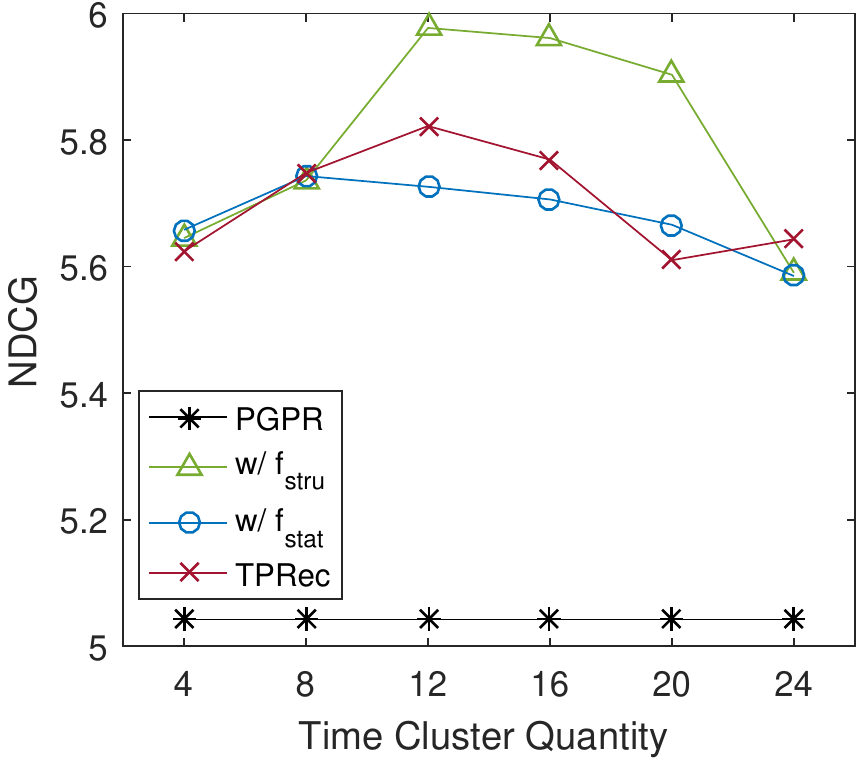}
		\end{minipage}
	}
	\subfloat[Recall (Cell).]{
		\begin{minipage}[t]{0.25\textwidth}
			\centering
			\includegraphics[width=1\textwidth]{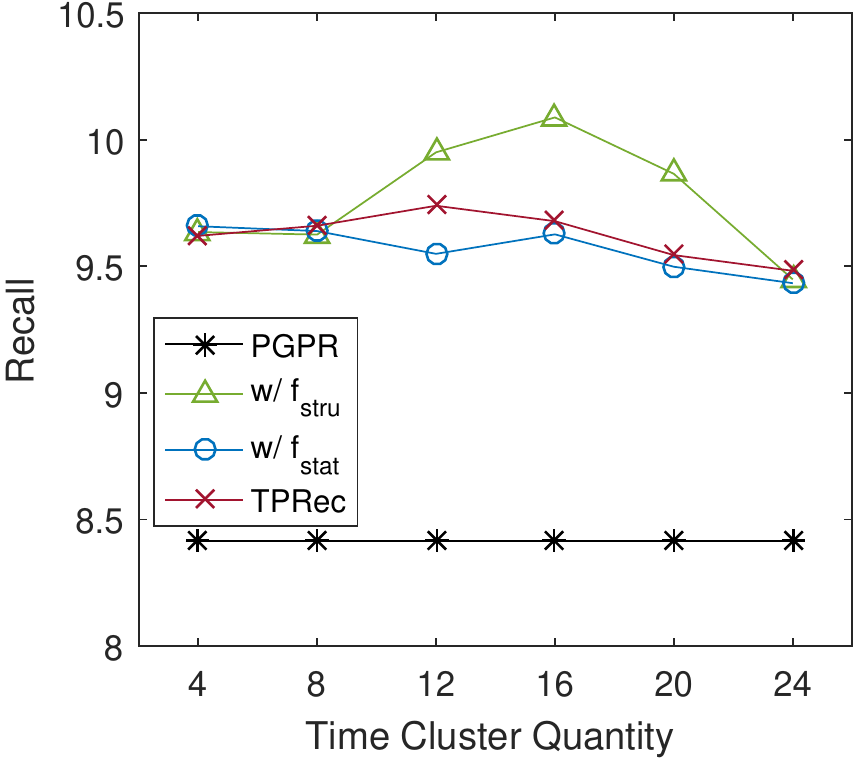}
		\end{minipage}
	}
	\subfloat[HR (Cell).]{
		\begin{minipage}[t]{0.25\textwidth}
			\centering
			\includegraphics[width=1\textwidth]{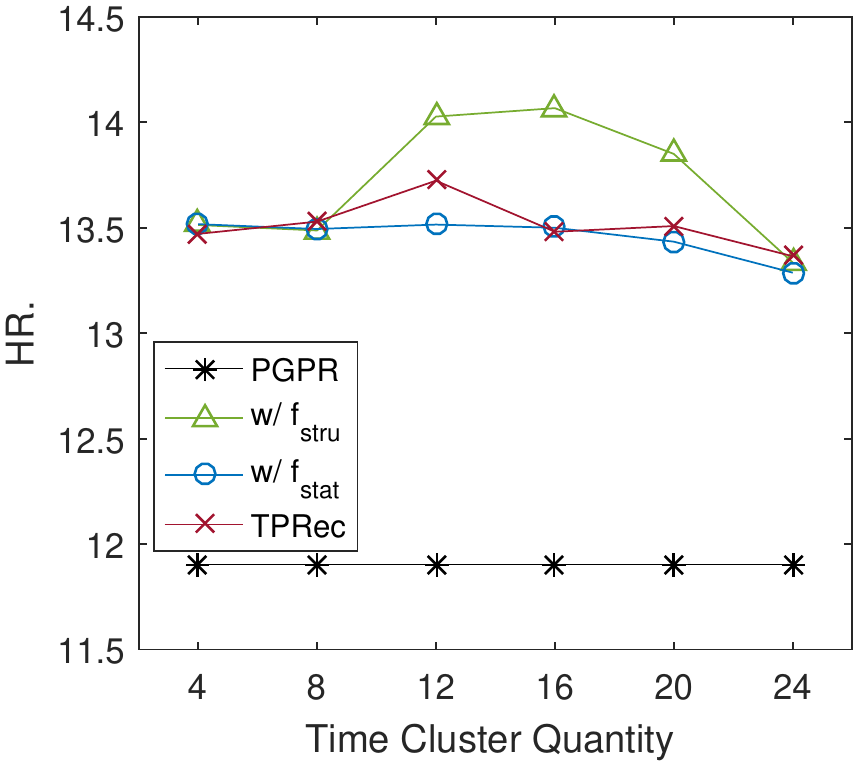}
		\end{minipage}
	}
	\subfloat[Precision (Cell).]{
		\begin{minipage}[t]{0.25\textwidth}
			\centering
			\includegraphics[width=1\textwidth]{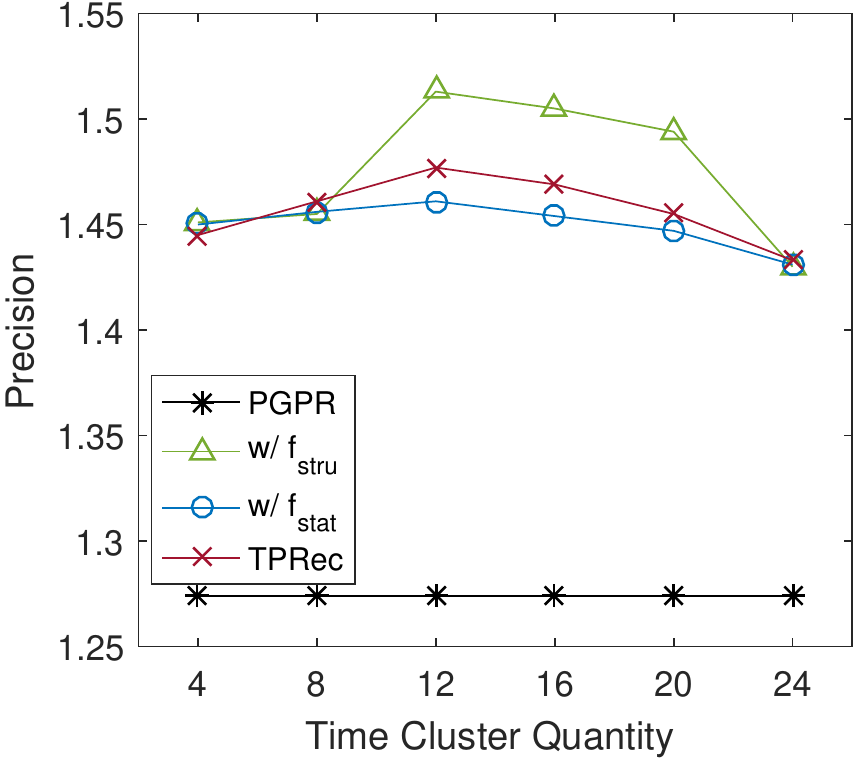}
		\end{minipage}
	}
	
	\subfloat[NDCG (Beauty).]{
		\begin{minipage}[t]{0.25\textwidth}
			\centering
			\includegraphics[width=1\textwidth]{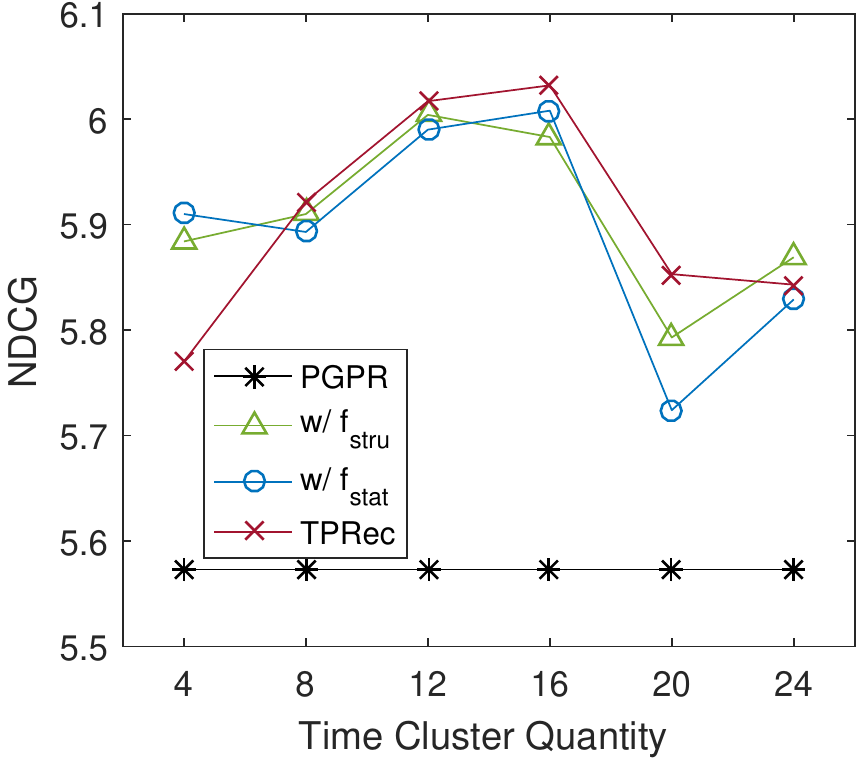}
		\end{minipage}
	}
		\subfloat[Recall (Beauty).]{
		\begin{minipage}[t]{0.25\textwidth}
			\centering
			\includegraphics[width=1\textwidth]{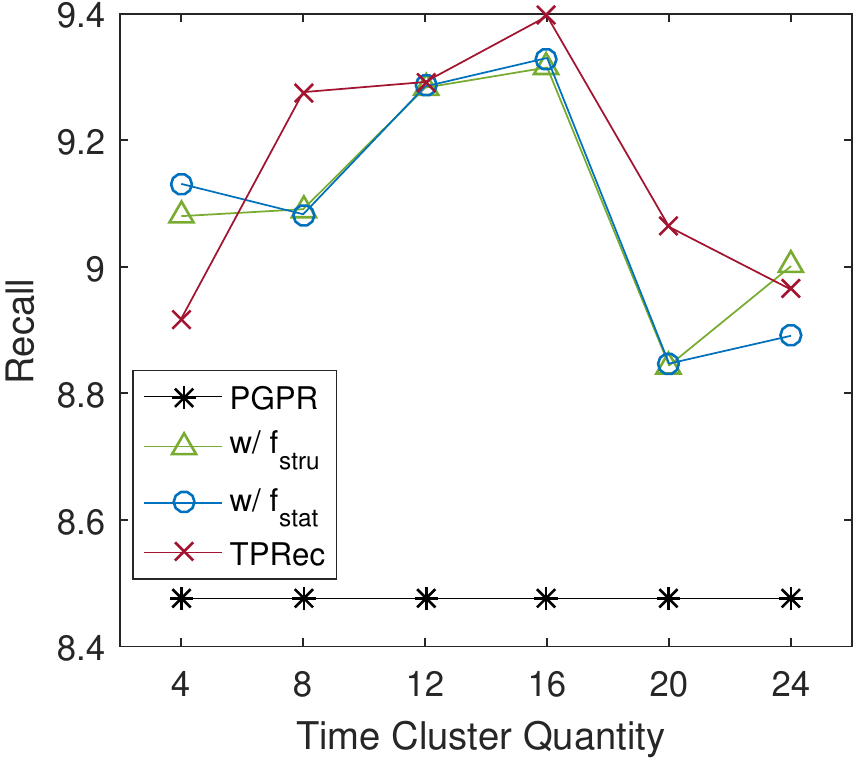}
		\end{minipage}
	}
	\subfloat[HR (Beauty).]{
		\begin{minipage}[t]{0.25\textwidth}
			\centering
			\includegraphics[width=1\textwidth]{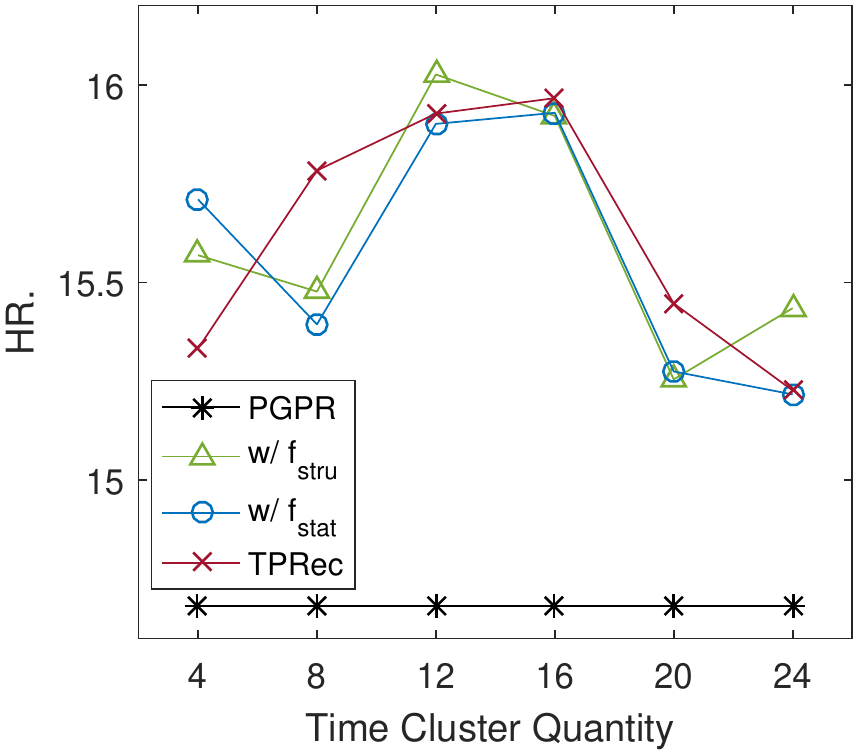}
		\end{minipage}
	}
	\subfloat[Precision (Beauty).]{
		\begin{minipage}[t]{0.25\textwidth}
			\centering
			\includegraphics[width=1\textwidth]{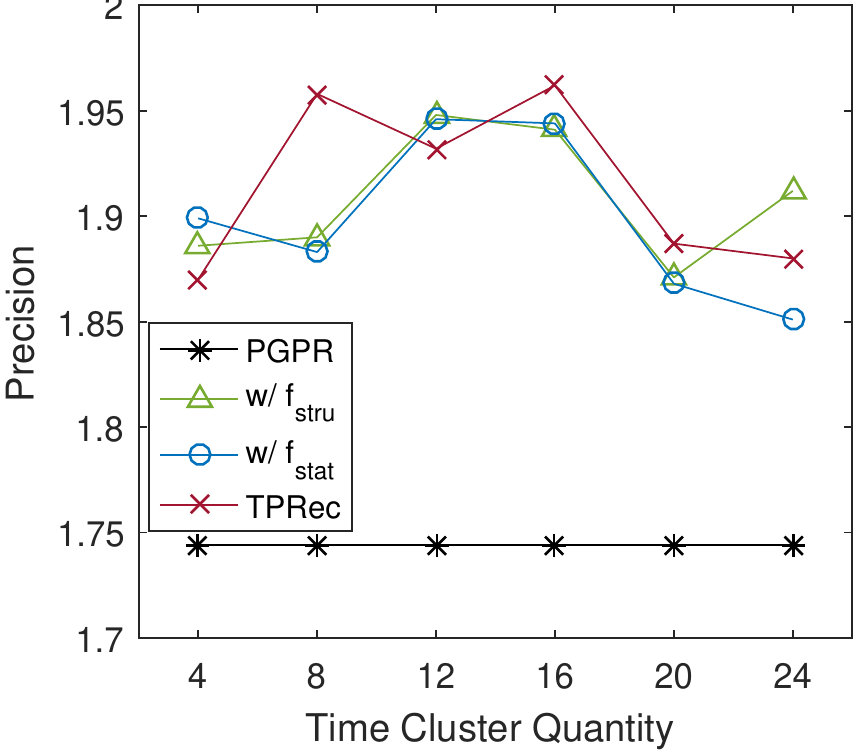}
		\end{minipage}
	}
\caption{ Effect of varying time cluster quantity $L$.
}
	\label{cluster_num}
\end{figure}

Figure \ref{cluster_num} presents the performance of four compared methods with different time 
cluster $L$ in range \{4, 8, 12, 16, 20, 24 \}. We make the following three observations: (1) With few exceptions, our \tkgrec\ consistently outperforms PGPR in all three datasets. 
(2) In the Cell Phones and Beauty dataset, all three versions of \tkgrec\ perform better when the number of time cluster $L = 12$ or $16$. And when $L$ is too small or too large, the performance of all three methods drops slightly.  
(3) In the Cloth dataset, when $L$ increases, the performance of our methods is relatively stable.
This is consistent with our strategy to adaptively determine time cluster number by BIC value (in Section \ref{parasetting}), and shows that BIC value can provide a suitable reference for clustering number $L$. 
%
%

\subsubsection{\textbf{Effect of Action Space Size.}}
\begin{figure}[]
	\centering
	\subfloat[NDCG (Clothing).]{
		\begin{minipage}[t]{0.25\textwidth}
			\centering
			\includegraphics[width=1\textwidth]{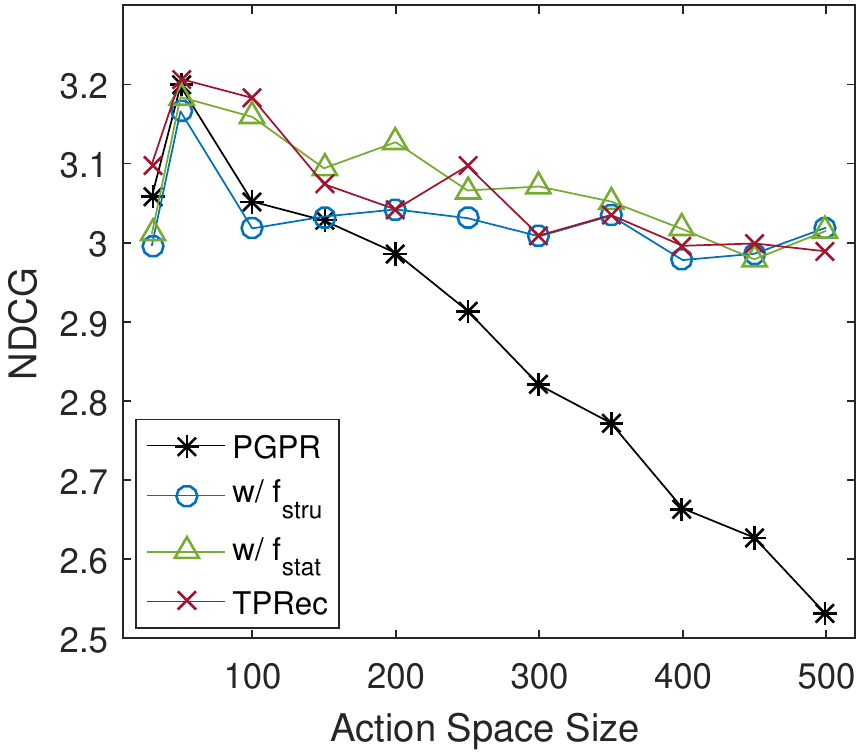}
		\end{minipage}
	}
	\subfloat[Recall (Clothing).]{
		\begin{minipage}[t]{0.25\textwidth}
			\centering
			\includegraphics[width=1\textwidth]{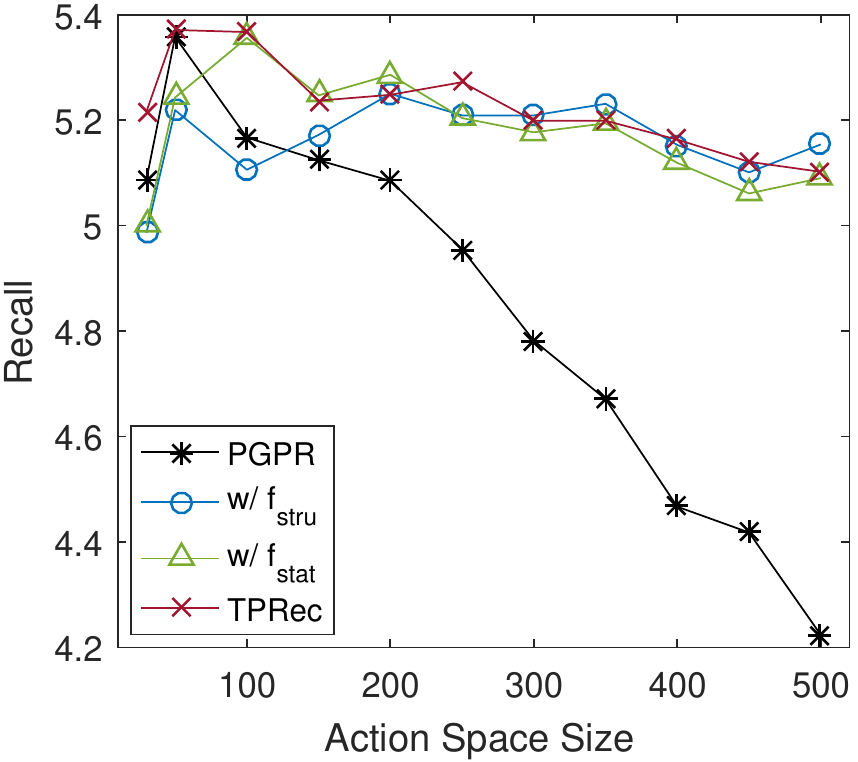}
		\end{minipage}
	}
	\subfloat[HR (Clothing).]{
		\begin{minipage}[t]{0.25\textwidth}
			\centering
			\includegraphics[width=1\textwidth]{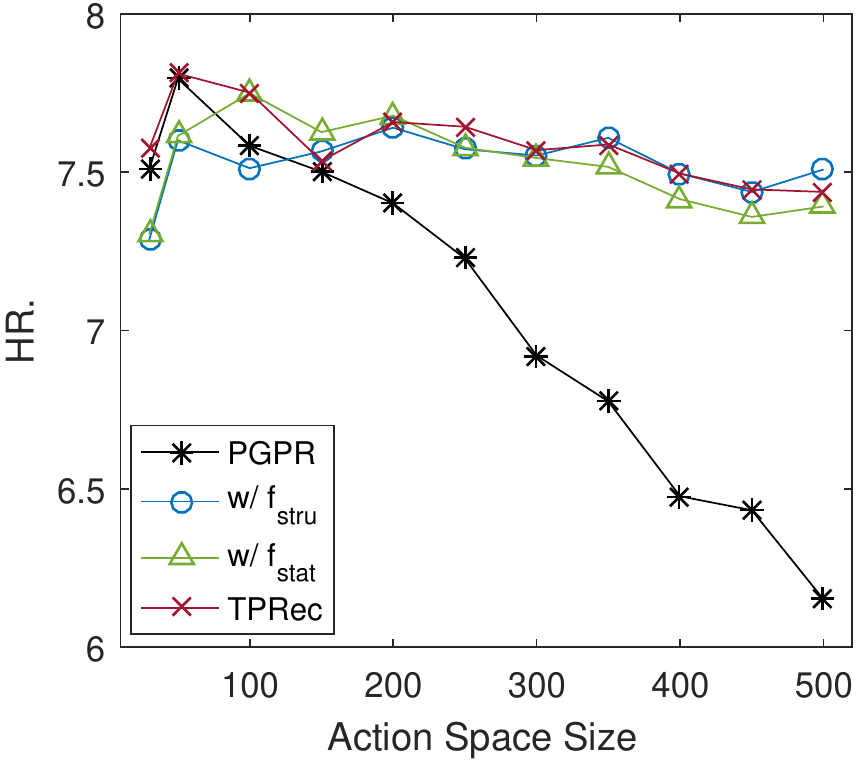}
		\end{minipage}
	}
	\subfloat[Precision (Clothing).]{
		\begin{minipage}[t]{0.25\textwidth}
			\centering
			\includegraphics[width=1\textwidth]{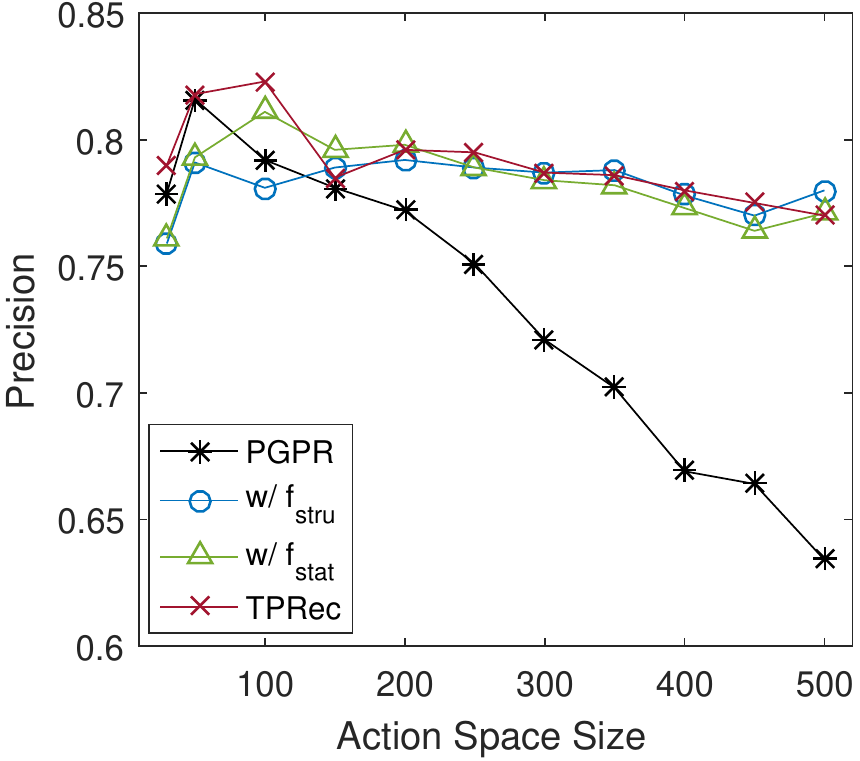}
		\end{minipage}
	}
	
	\subfloat[NDCG (Cell).]{
		\begin{minipage}[t]{0.25\textwidth}
			\centering
			\includegraphics[width=1\textwidth]{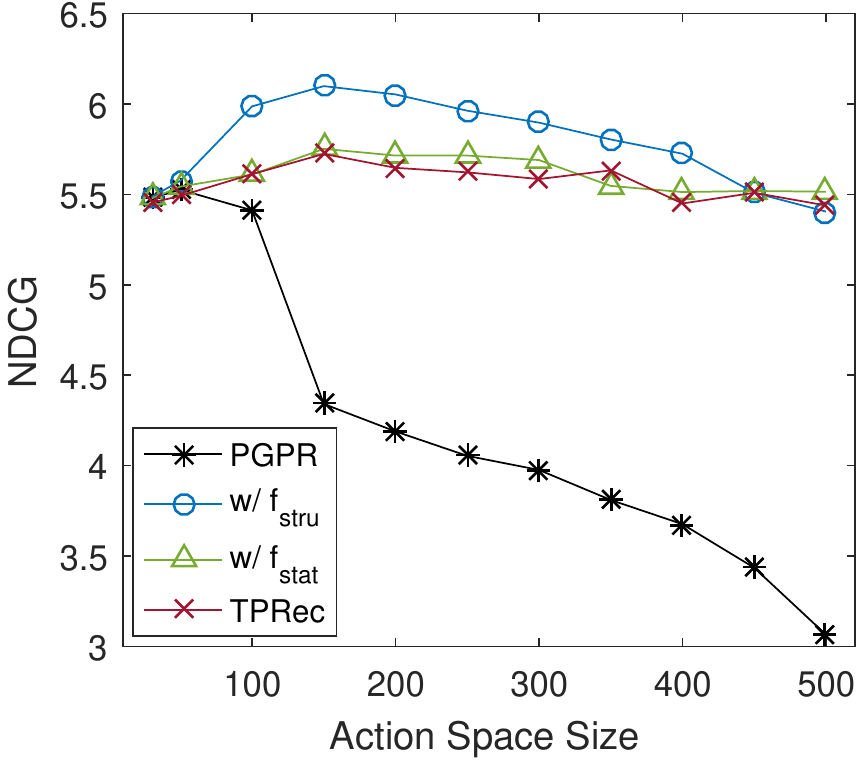}
		\end{minipage}
	}
	\subfloat[Recall (Cell).]{
		\begin{minipage}[t]{0.25\textwidth}
			\centering
			\includegraphics[width=1\textwidth]{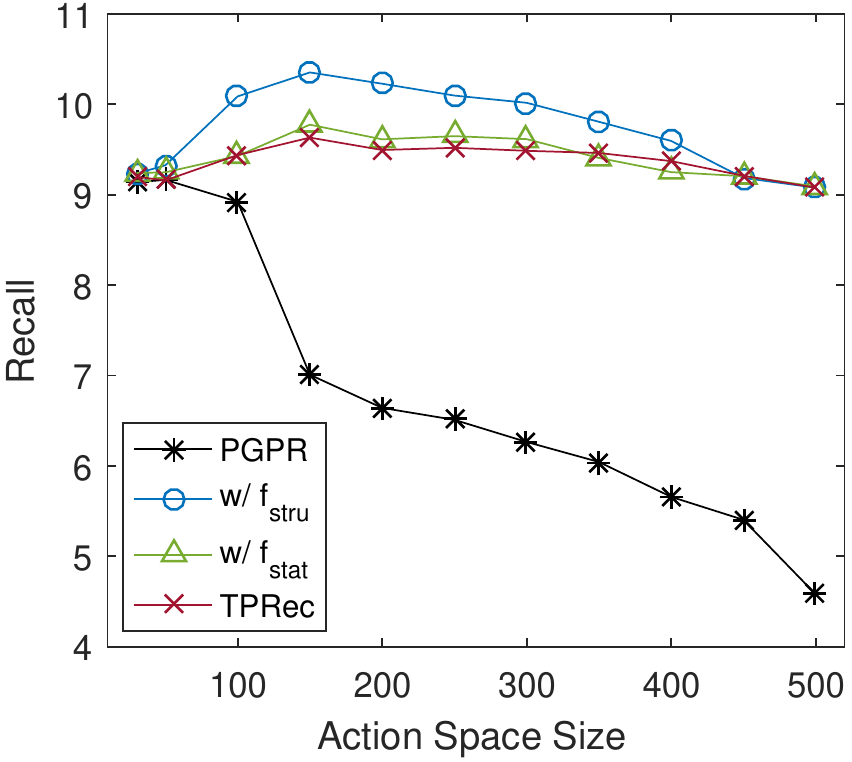}
		\end{minipage}
	}
	\subfloat[HR (Cell).]{
		\begin{minipage}[t]{0.25\textwidth}
			\centering
			\includegraphics[width=1\textwidth]{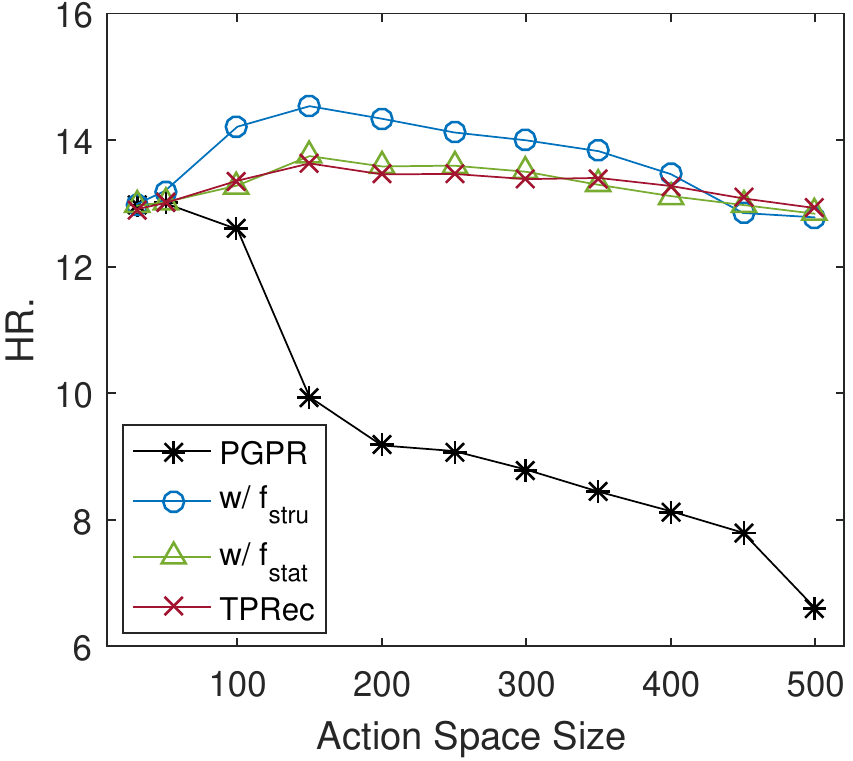}
		\end{minipage}
	}
	\subfloat[Precision (Cell).]{
		\begin{minipage}[t]{0.25\textwidth}
			\centering
			\includegraphics[width=1\textwidth]{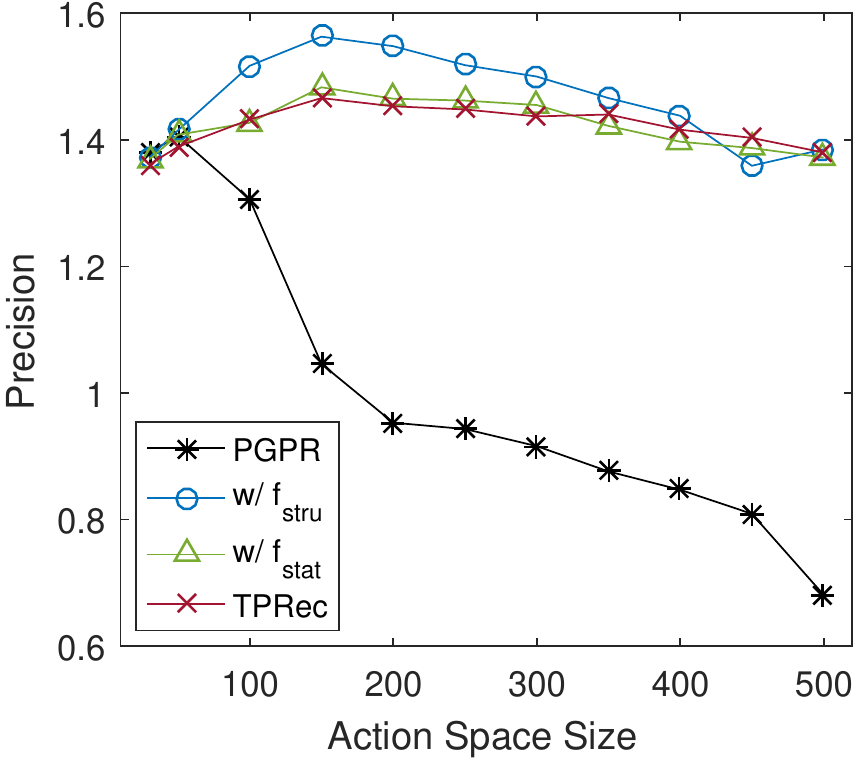}
		\end{minipage}
	}
	
	\subfloat[NDCG (Beauty).]{
		\begin{minipage}[t]{0.25\textwidth}
			\centering
			\includegraphics[width=1\textwidth]{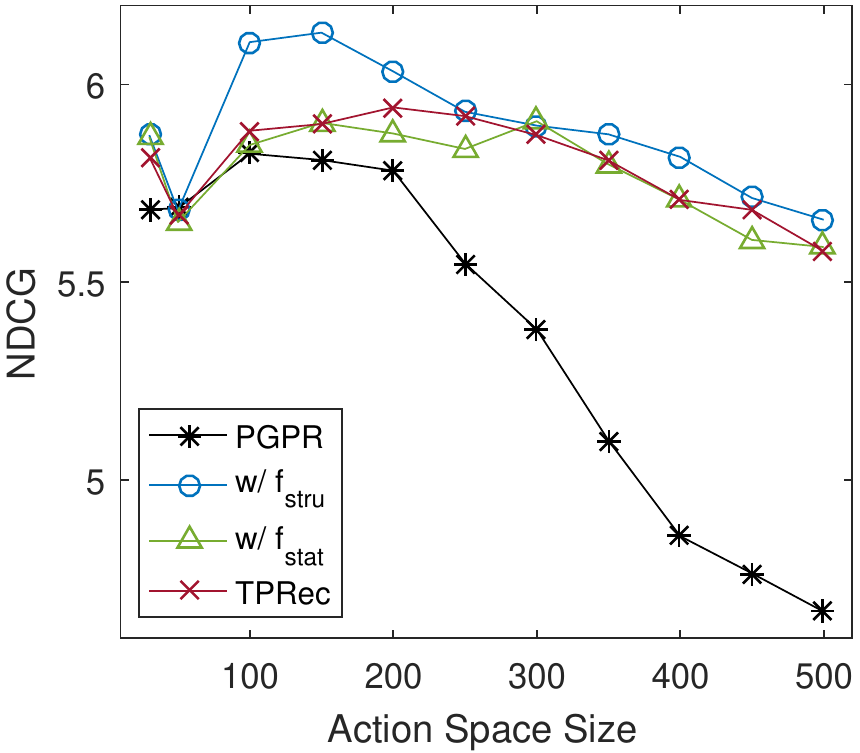}
		\end{minipage}
	}
		\subfloat[Recall (Beauty).]{
		\begin{minipage}[t]{0.25\textwidth}
			\centering
			\includegraphics[width=1\textwidth]{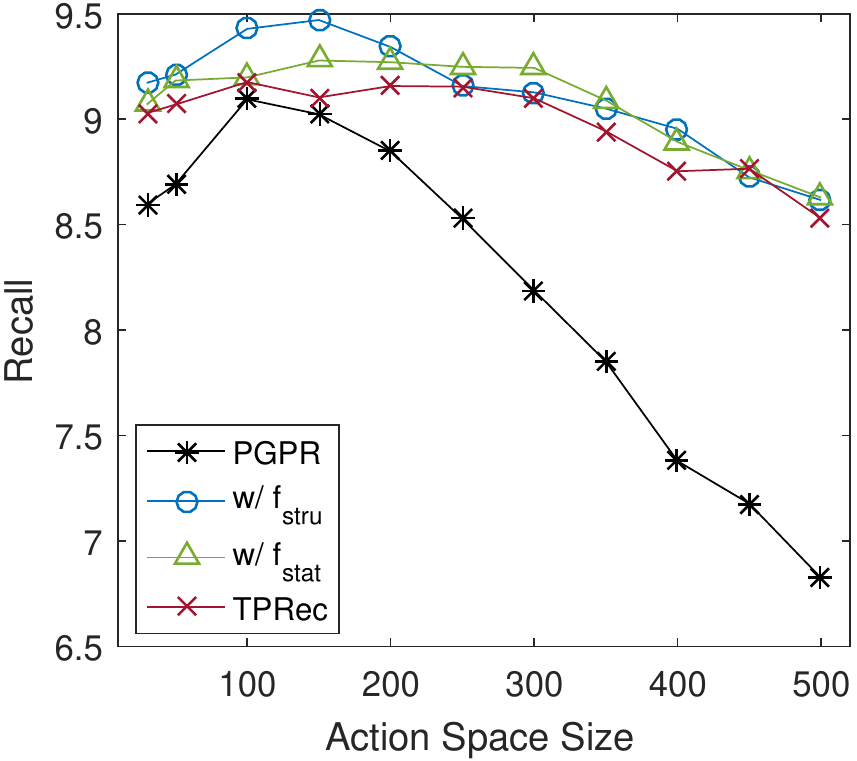}
		\end{minipage}
	}
	\subfloat[HR (Beauty).]{
		\begin{minipage}[t]{0.25\textwidth}
			\centering
			\includegraphics[width=1\textwidth]{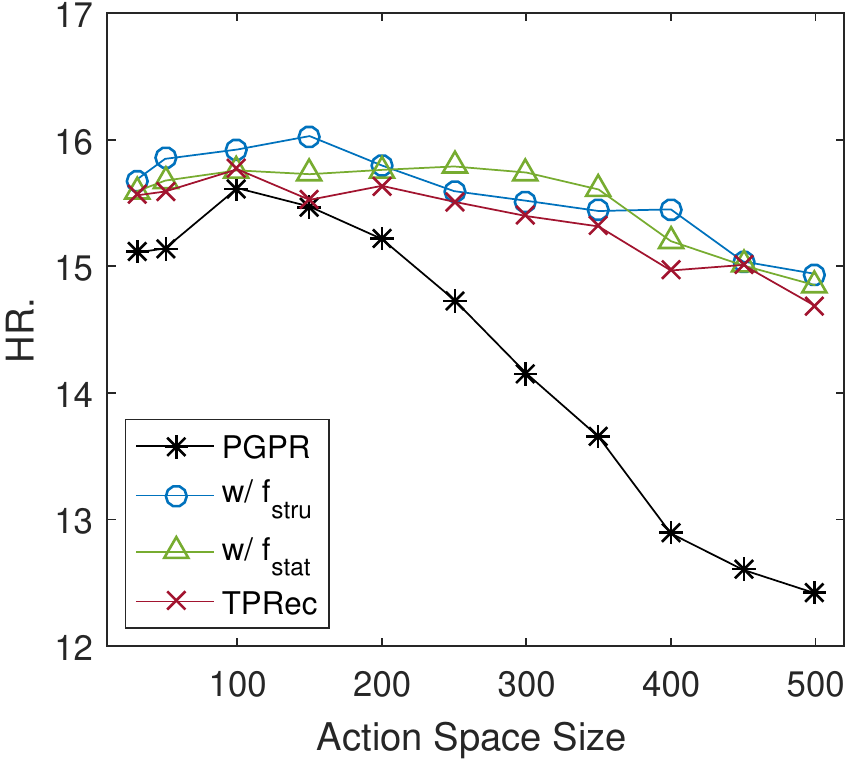}
		\end{minipage}
	}
	\subfloat[Precision (Beauty).]{
		\begin{minipage}[t]{0.25\textwidth}
			\centering
			\includegraphics[width=1\textwidth]{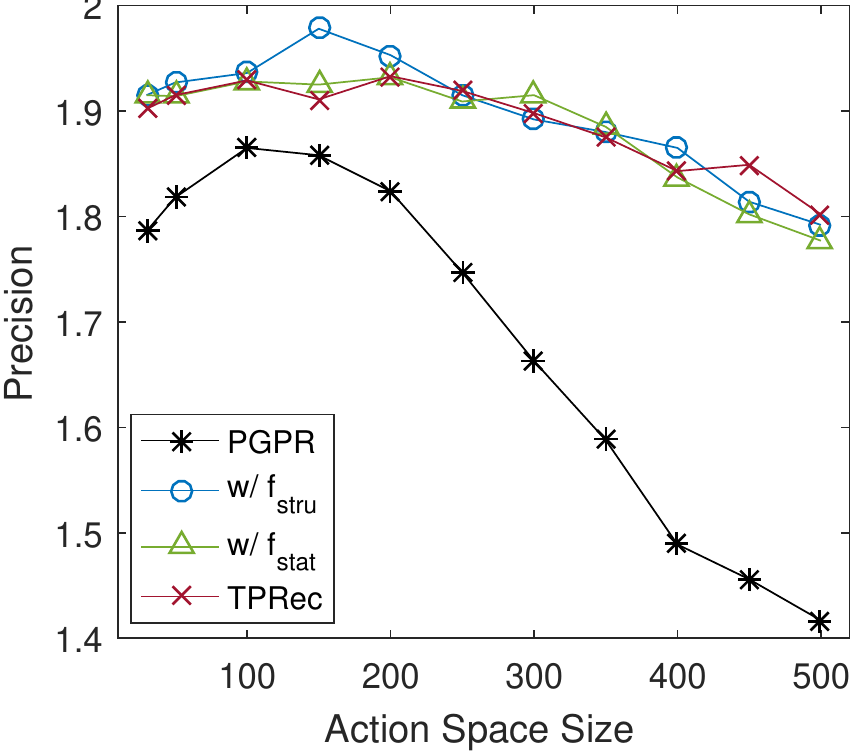}
		\end{minipage}
	}
\caption{ Effect of varying action space size $\epsilon$.
}
	\label{actionSpace}
\end{figure}

 Figure \ref{actionSpace} presents the performance of four compared methods with different action space size $\epsilon$ varying from 100 to 500. We make the following two observations: 
  (1) With few exceptions, all three versions of \tkgrec\ consistently outperform PGPR with various $\epsilon$, demonstrating the effectiveness of exploiting temporal information. 
(2) As the action space size $\epsilon$ increases, generally speaking, the performance will become better at the beginning.  The reason is that when the action space size $\epsilon$ is too small, the model is likely to miss the relevant paths. It also can be seen from Figure \ref{act_invalid}, that a smaller action space size (\eg\ $ \epsilon\leq 200$) would incur a larger number of invalid users.
 (3) But when the $\epsilon$ surpasses a certain threshold, the performance of all the methods drops with $\epsilon$ further increasing. The reason is that a larger action space would increase the risk of retrieving irrelevant paths. Nonetheless, our \tkgrec\ drops relatively mildly. It demonstrates our \tkgrec\ is more robust to $\epsilon$ and able to identify high-quality paths by leveraging temporal information.

\subsubsection{\textbf{Effect of State History Length.}}
\begin{figure}[]
	\centering
	\subfloat[NDCG (Clothing).]{
		\begin{minipage}[t]{0.25\textwidth}
			\centering
			\includegraphics[width=1\textwidth]{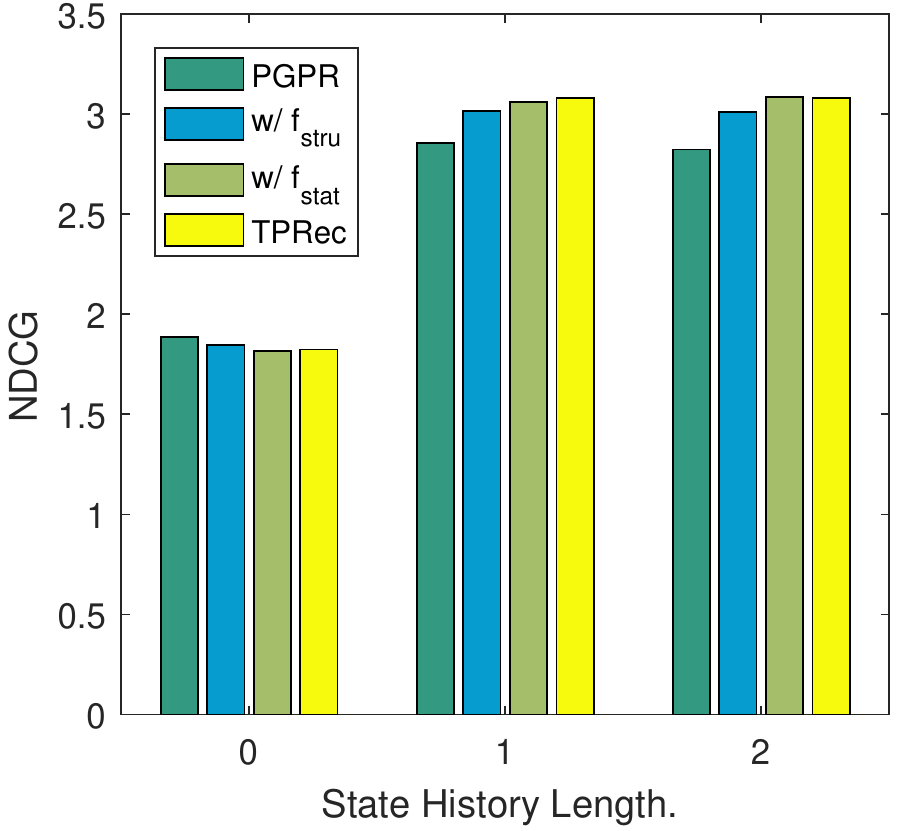}
		\end{minipage}
	}
	\subfloat[Recall (Clothing).]{
		\begin{minipage}[t]{0.25\textwidth}
			\centering
			\includegraphics[width=1\textwidth]{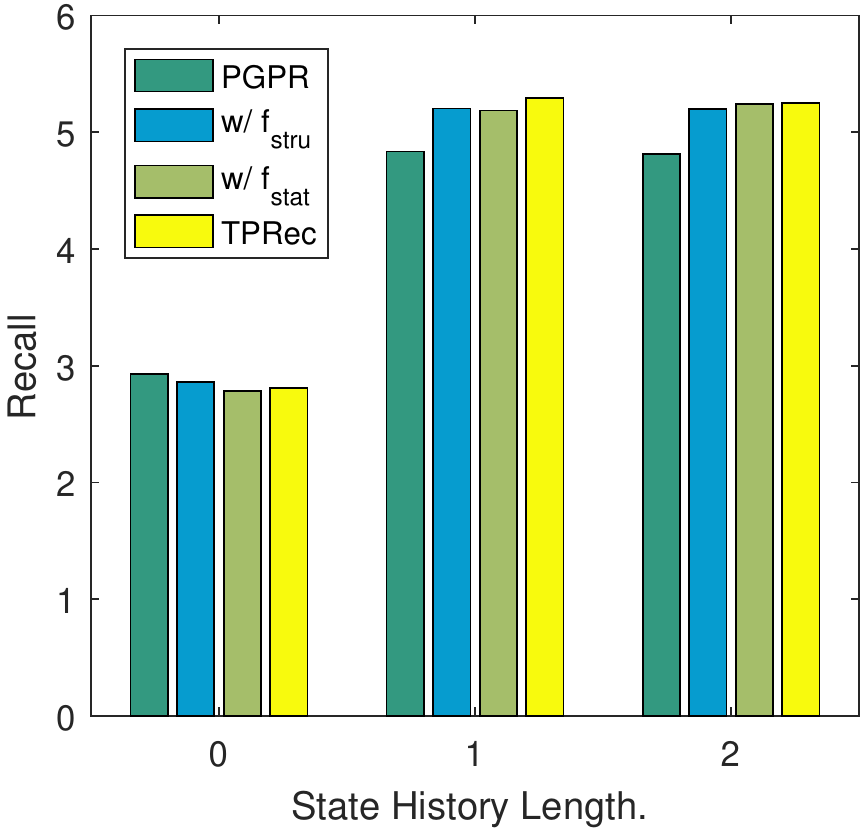}
		\end{minipage}
	}
	\subfloat[HR (Clothing).]{
		\begin{minipage}[t]{0.25\textwidth}
			\centering
			\includegraphics[width=1\textwidth]{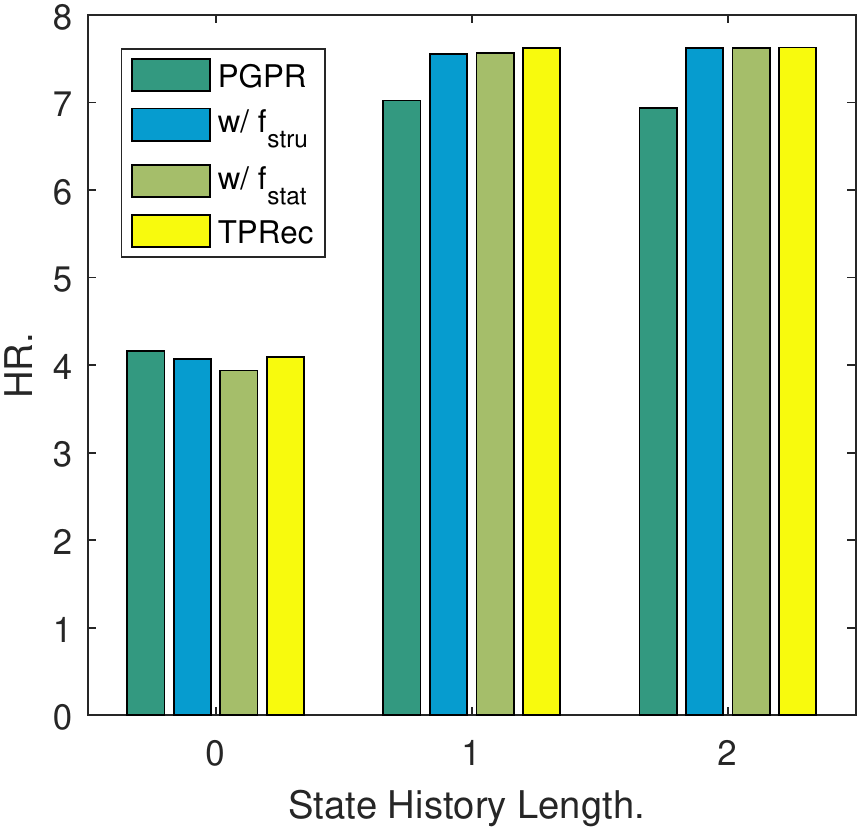}
		\end{minipage}
	}
	\subfloat[Precision (Clothing).]{
		\begin{minipage}[t]{0.25\textwidth}
			\centering
			\includegraphics[width=1\textwidth]{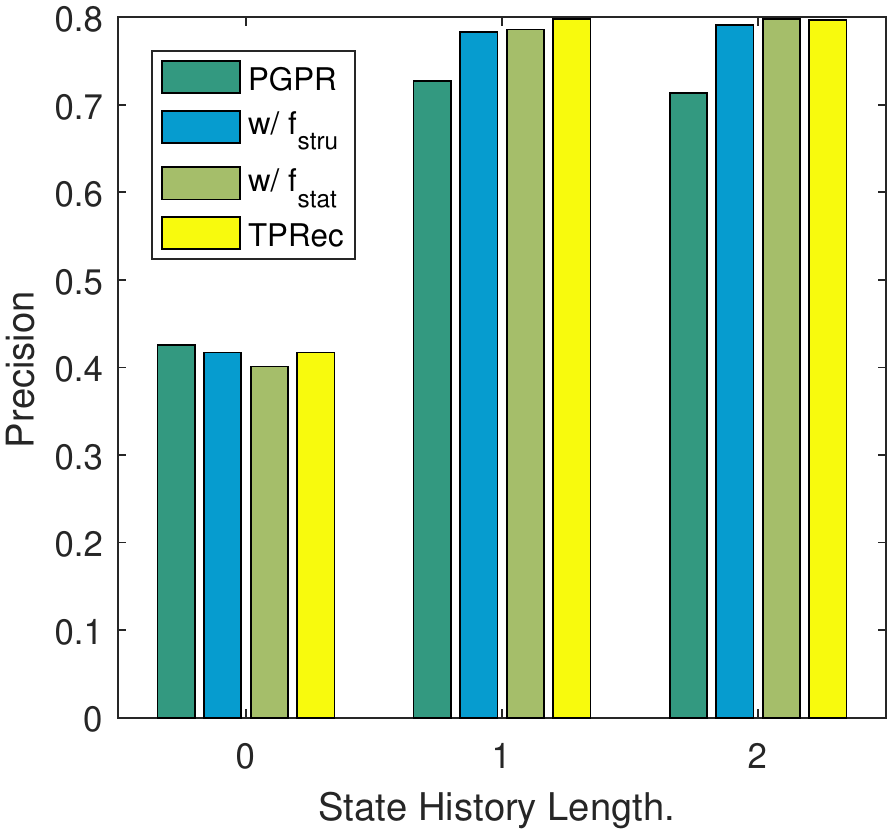}
		\end{minipage}
	}
	
	\subfloat[NDCG (Clothing).]{
		\begin{minipage}[t]{0.25\textwidth}
			\centering
			\includegraphics[width=1\textwidth]{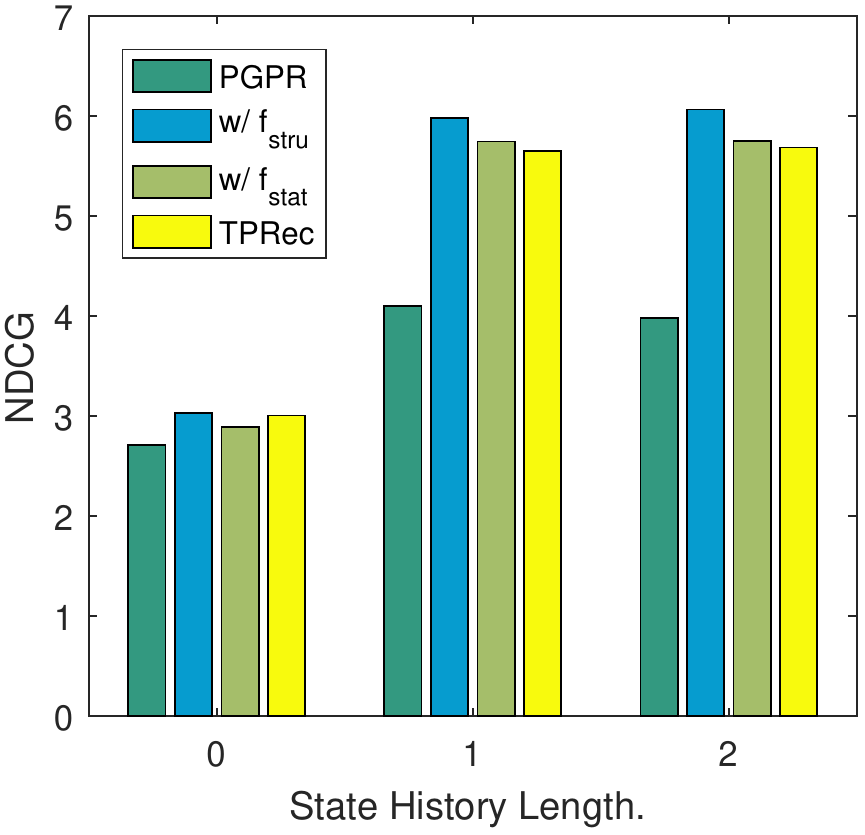}
		\end{minipage}
	}
	\subfloat[Recall (Cell).]{
		\begin{minipage}[t]{0.25\textwidth}
			\centering
			\includegraphics[width=1\textwidth]{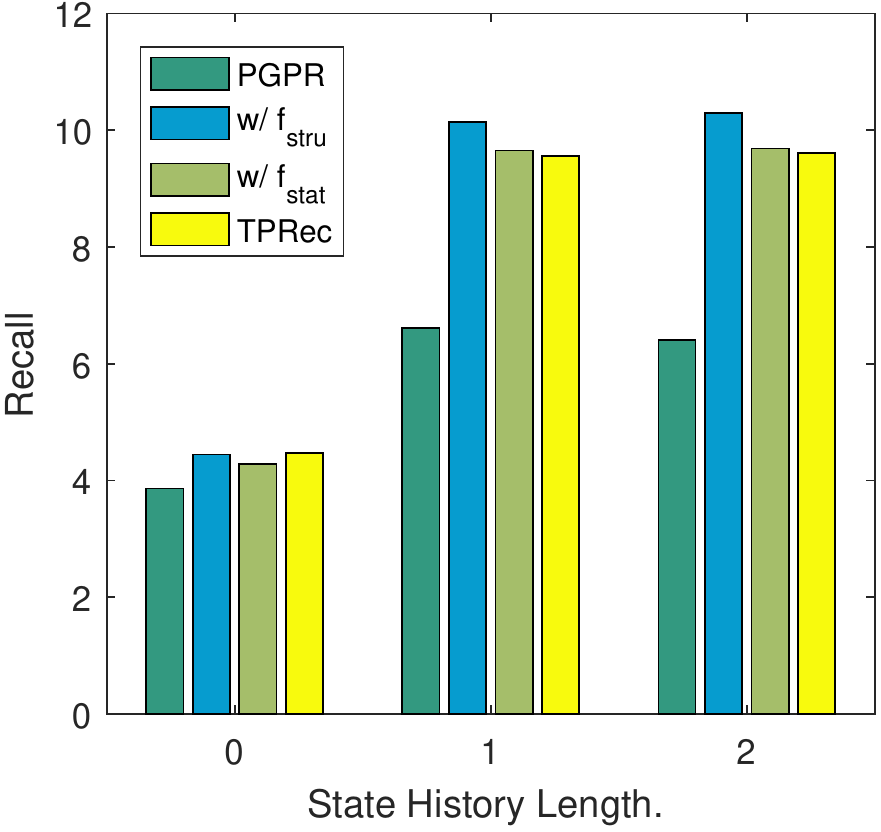}
		\end{minipage}
	}
	\subfloat[HR (Cell).]{
		\begin{minipage}[t]{0.25\textwidth}
			\centering
			\includegraphics[width=1\textwidth]{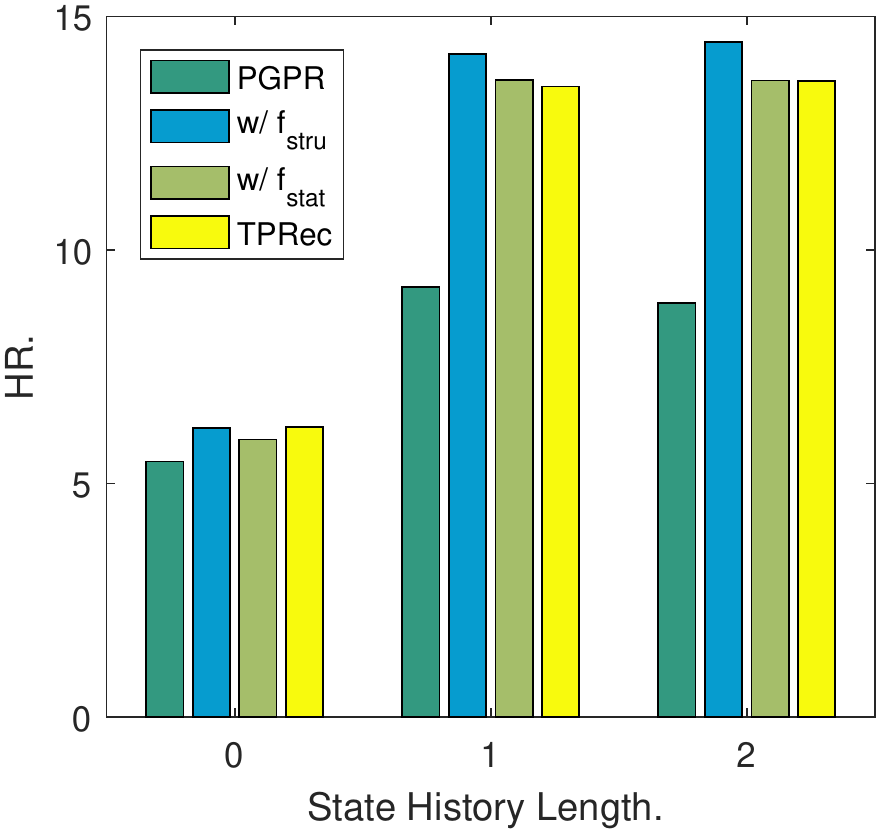}
		\end{minipage}
	}
	\subfloat[Precision (Cell).]{
		\begin{minipage}[t]{0.25\textwidth}
			\centering
			\includegraphics[width=1\textwidth]{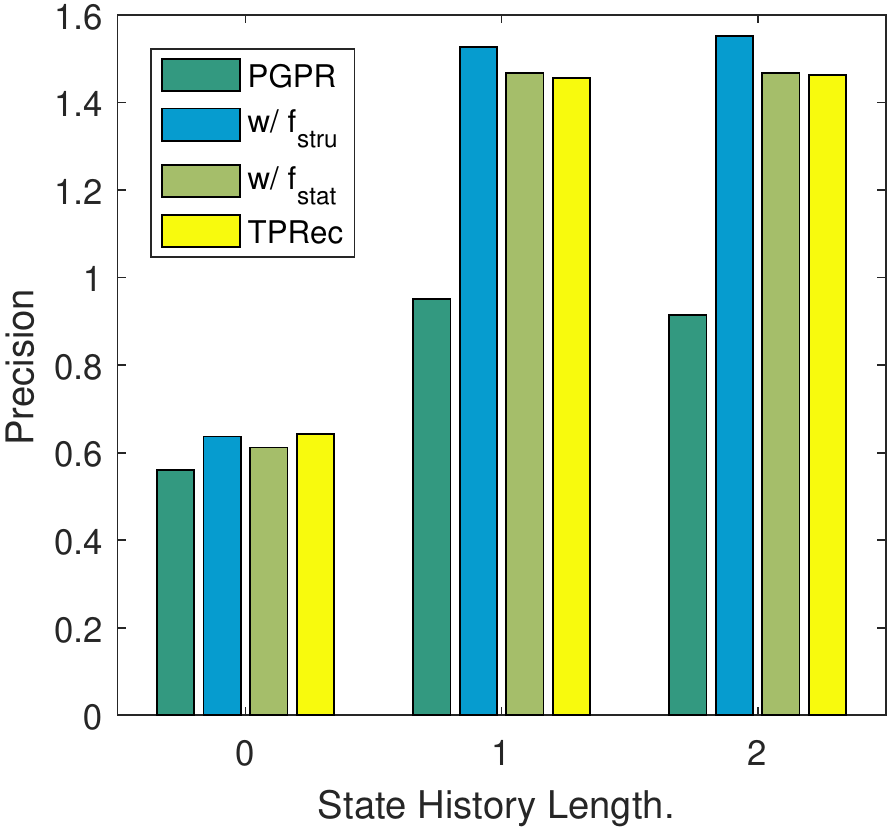}
		\end{minipage}
	}
	
	\subfloat[NDCG (Beauty).]{
		\begin{minipage}[t]{0.25\textwidth}
			\centering
			\includegraphics[width=1\textwidth]{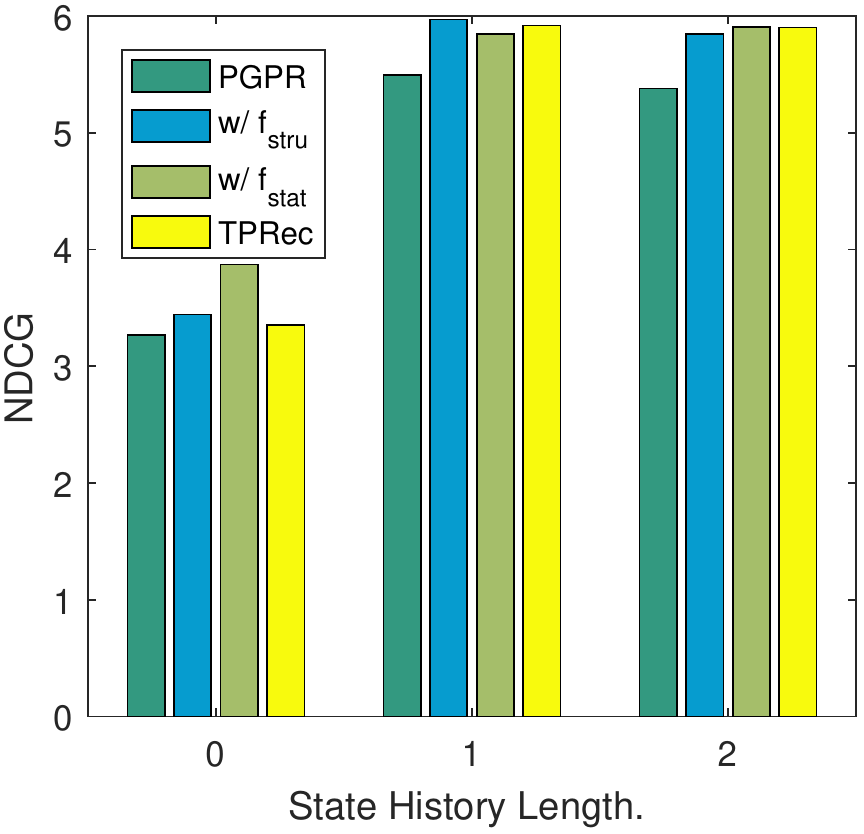}
		\end{minipage}
	}
		\subfloat[Recall (Beauty).]{
		\begin{minipage}[t]{0.25\textwidth}
			\centering
			\includegraphics[width=1\textwidth]{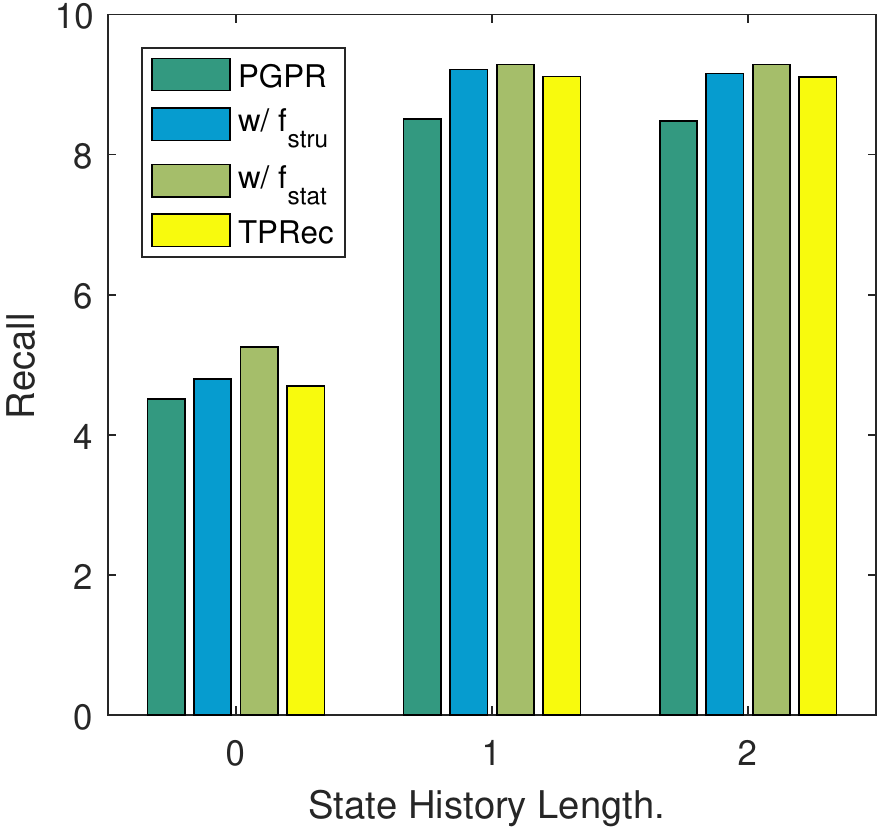}
		\end{minipage}
	}
	\subfloat[HR (Beauty).]{
		\begin{minipage}[t]{0.25\textwidth}
			\centering
			\includegraphics[width=1\textwidth]{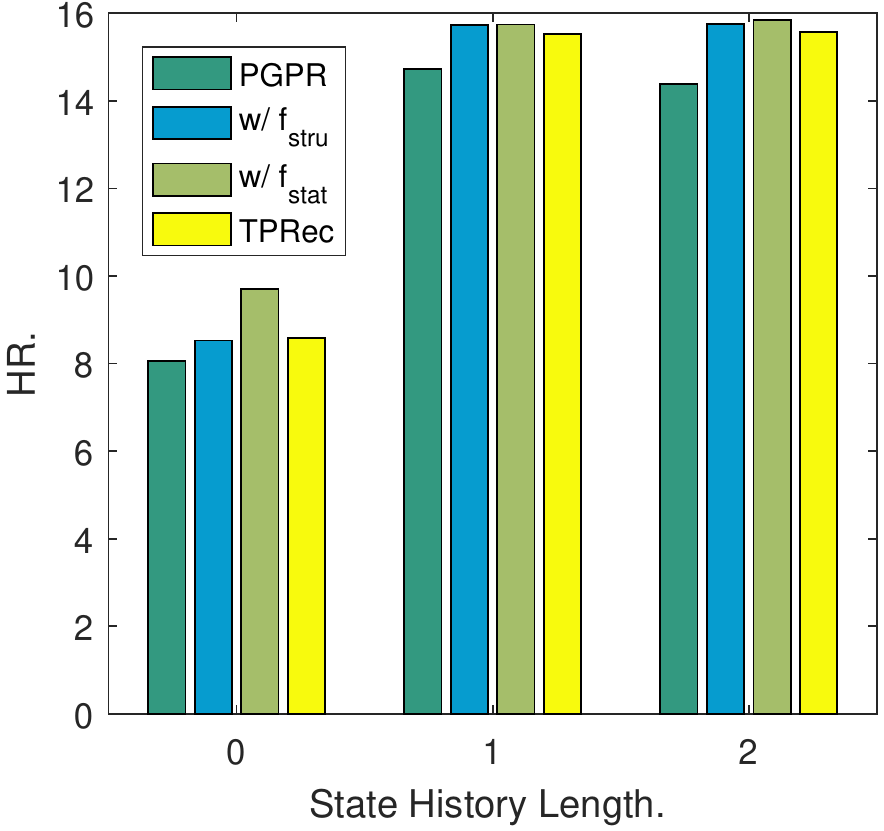}
		\end{minipage}
	}
	\subfloat[Precision (Beauty).]{
		\begin{minipage}[t]{0.25\textwidth}
			\centering
			\includegraphics[width=1\textwidth]{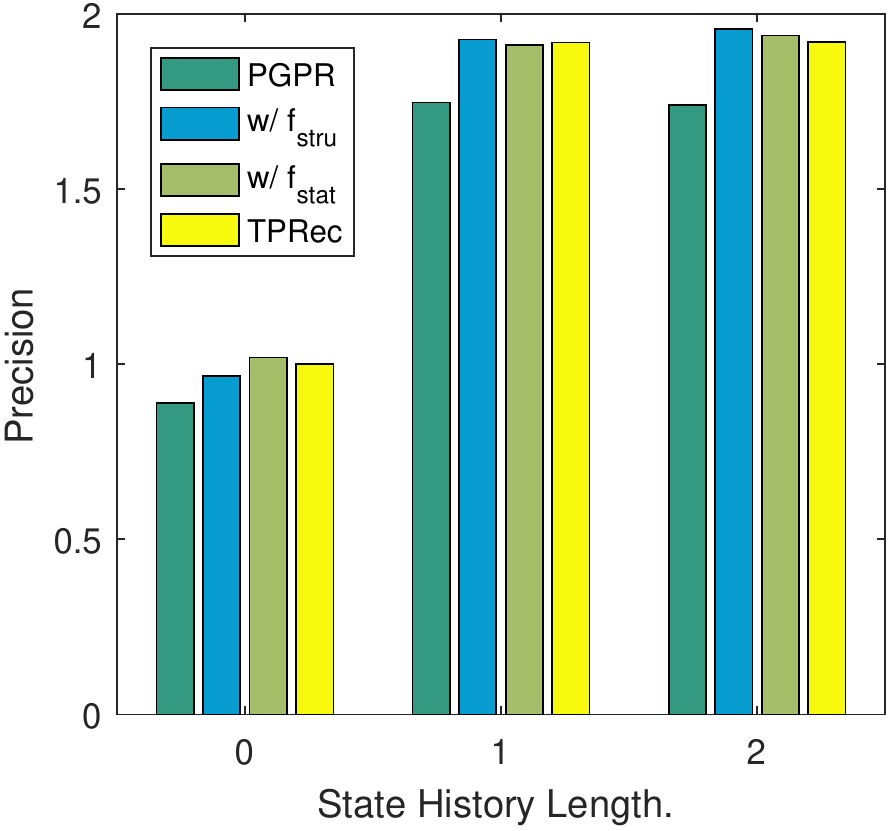}
		\end{minipage}
	}
\caption{ Effect of varying state history length $k'$.
}
	\label{stateHistory}
\end{figure}

Figure \ref{stateHistory} presents the performance of four compared methods with different State History Length $k'$ in range $\{0,1,2\}$. We make the following observations: (1) with few exceptions, our \tkgrec\ consistently outperforms PGPR. (2) For all methods, the versions with encoding 1-step or 2-step history consistently outperform those without encoding history. This result validates the effectiveness of exploiting historical states to learn policy. But, we observe that encoding 2-step history may not be superior over 1-step. Too long history will bring some irrelevant information or even noises, which deteriorates recommendation performance.

\subsection{Evaluating Explanations (RQ4)}
How to evaluate explanations in recommender system is a hard problem. Fortunately, inspired by \cite{countER, ADAC}, we mathematically define a user-oriented evaluation metric to evaluate explainability. And finally, we give a case study to demonstrate explanations.

\subsubsection{\textbf{Explainability Evaluation}} 
To automatically evaluate the quality of explanations generated by \tkgrec, we adopt the user's reviews on items as the ground truth reason about why the user purchased the item, as previous works\cite{countER, ADAC, chen2019co, chen2018neural, li2020generate, wang2018explainable, DIR,  wang2021towards, igv} do. Here, for reasoning paths $\tau_{u,\hat{V}}$ that an item $\hat{v} \in \hat{V}$ recommended to user $u$ before position $K$, we extract all the \textit{Word} feature entities as an explanation list, which is defined as $S_{u} = [s^{(0)}_{u}, s^{(1)}_{u}, ..., s^{(r)}_{u}]$. Thus, if a reasoning path contains more entities that are mentioned in ground truth reviews, it will achieve better explainability. More specifically, we filter less salient review words if their frequency is more than 5000 and TF-IDF < 0.1, 
and the remaining words are our ground truth reason $G_u = [g^{(0)}_{u}, g^{(1)}_{u}, ..., g^{(n)}_{u}]$.

Then for each user, we calculate the $Recall$, $Precision$ and $F_1$ score of the reasoning path's $Word$ entities $S_{u}$ with regard to the ground truth $G_u$:

\begin{align}   \label{metric_explain}
	 Recall &= \frac{|S_{u} \cap G_u|}{|G_{u}| + 1}, & Precision &= \frac{|S_{u} \cap G_u|}{|S_{u}| + 1}, &  F_1 &=  2 \cdot \frac{Precision \cdot Recall}{Precision + Recall + 1},
\end{align}
where $Recall$ measures the percentage of aspects liked by the user included in our explanation, $Precision$ indicates how much our explanation is liked by the user, and $F_1$ score is the harmonic mean between those two. And the $+ 1$ in the denominator is used to avoid dividing by $0$. Finally, we average the scores of all users to evaluate explainability.

\begin{table}[t!]
\caption{Explanation performance comparison between PGPR and \tkgrec\ on different datasets.}\label{explainCmp}
\begin{tabular}{c|crr|crr|crr}
\hline \hline
\textbf{Dataset} & \multicolumn{3}{c|}{\textbf{Cloth}}                                                                & \multicolumn{3}{c|}{\textbf{Cellphone}}                                                            & \multicolumn{3}{c}{\textbf{Beauty}}                                                               \\ \hline
\textbf{Metric}  & \textbf{Recall}            & \multicolumn{1}{c}{\textbf{Prec.}} & \multicolumn{1}{c|}{\textbf{F1}} & \textbf{Recall}            & \multicolumn{1}{c}{\textbf{Prec.}} & \multicolumn{1}{c|}{\textbf{F1}} & \textbf{Recall}            & \multicolumn{1}{c}{\textbf{Prec.}} & \multicolumn{1}{c}{\textbf{F1}} \\ \hline
\textbf{PGPR}    & \multicolumn{1}{r}{23.771} & 90.750                             & 19.340                           & \multicolumn{1}{r}{22.970} & 90.623                             & 18.522                           & \multicolumn{1}{r}{19.296} & 90.879                             & 15.936                          \\
\textbf{\tkgrec}   & \multicolumn{1}{r}{24.642} & 90.871                             & 21.133                           & \multicolumn{1}{r}{23.637} & 90.895                             & 19.043                           & \multicolumn{1}{r}{20.093} & 90.891                    & 17.471                          \\ \hline \hline
\end{tabular}
\end{table}

From Table \ref{explainCmp}, our \tkgrec\ outperforms PGPR in all three datasets. Though not strictly, it provides an angle to show that the use of temporal information can achieve better explainability.

\subsubsection{\textbf{Case Study}}
\begin{figure}[]
\includegraphics[width=\textwidth]{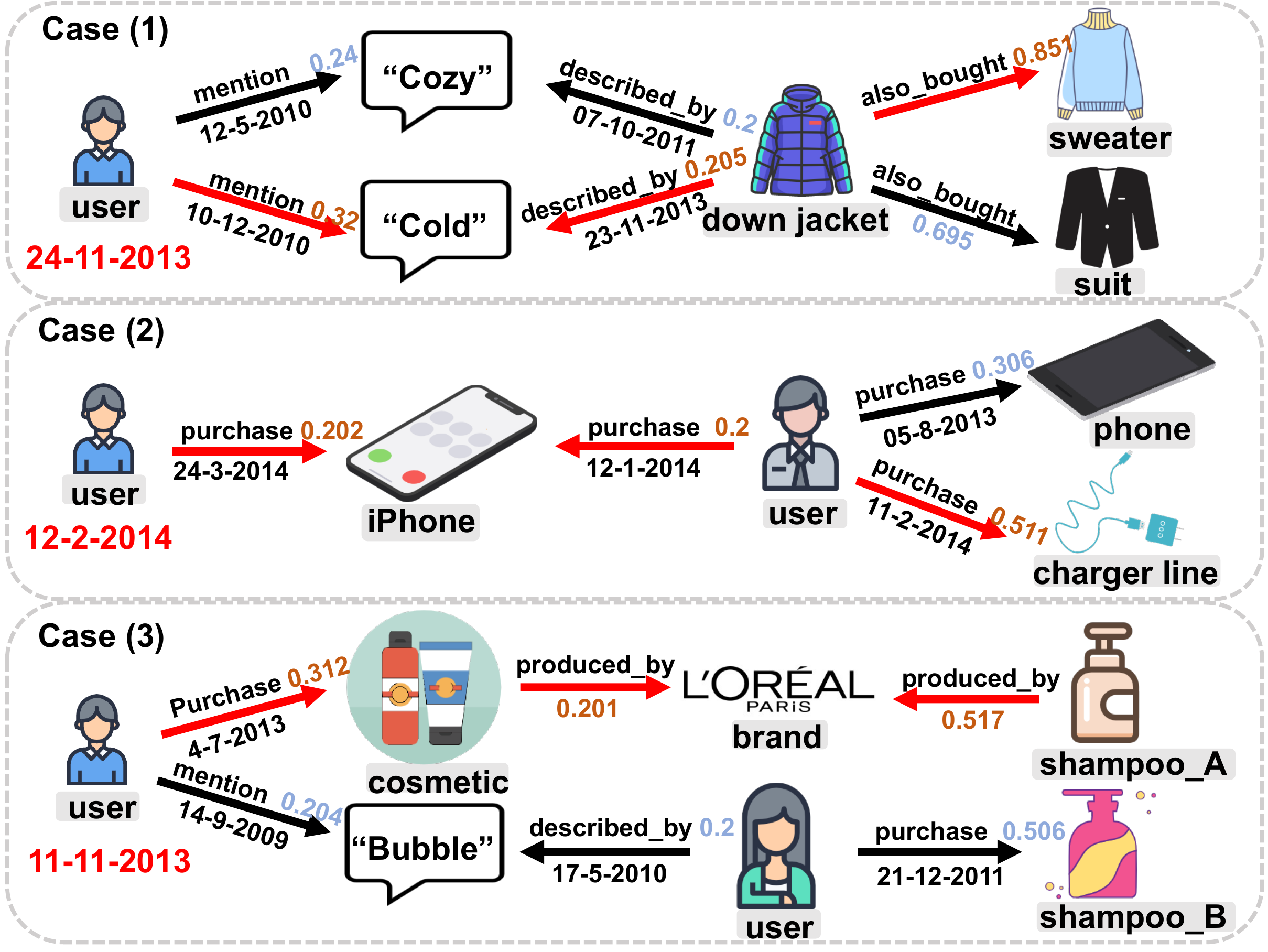}
\caption{Case study on how temporal information benefits explainable
recommendation. The timestamp in red font below ``user'' depicts the recommended
date. The path marked in red denotes the top relevant one predicted by TPRec. }
\label{caseStudy}
\end{figure}

To better understand how temporal information boosts the quality of explainable recommendation, we conduct a case study based on the result of our \tkgrec. As shown in Figure \ref{caseStudy}, we provide several examples of the learned reasoning paths with hop scores, where the path marked in red denotes the top relevant one predicted by \tkgrec.


The first example (Case 1) comes from the \textit{Clothing} dataset, where we make a recommendation for a specific user on 24-11-2013. By leveraging temporal information, our \tkgrec~ can deduce the weather is winter and correspondingly identify the most relevant word is ``cold''. Based on such knowledge, \tkgrec~ is able to retrieve more relevant ``sweater'' to the user, while PGPR that does not use temporal information would fall short. In the second example (Case 2), there are two users who both bought an ``iPhone'' on 24-3-2014 and 12-1-2014 respectively. The second user also purchased a "phone" on 05-8-2013 and a "charger line" on 11-2-2014. As we can see, our \tkgrec~ recommends ``charge line'' instead of ``phone'' as it could capture the common sense that a person is unlikely to buy two mobile phones in a short time. The last example (Case 3) depicts two reasoning paths of a user purchasing "shampoo". Our \tkgrec~ could differentiate the better one --- the red path is more reasonable, as the link ``purchase'' happened on the day 4-7-2013, which is the shopping festival as the day 11-11-2013. Naturally, the user on 11-11-2013 is more likely to have similar needs as on 4-7-2013, and the item from the red path is more likely to be favored by the user.

%% file: 5-RelatedWork.tex
\section{Related Work}
\label{sec:related}

We divide the related works into three categories: (\romannumeral1) knowledge 
graph based recommendation, (\romannumeral2) temporal knowledge graph embedding
and (\romannumeral3)  temporal information in 
recommendation.

\subsection{Knowledge Graph Based Recommendation}

Existing studies on KG-based recommendation systems \cite{surveyRec} can be roughly grouped into two sub-categories: embedding-based methods and path-based
methods.

Embedding-based methods adopt a common general model, where the latent vector
$\boldsymbol{u}_i$ of each user $u_i$ and item embedding $\boldsymbol{v}_j$ can 
be
extracted from KGE, and the probability  of item $v_j$ being recommended to
user $u_i$ can be calculated by $\hat{y}_{i, j}=f\left(\mathbf{u}_{i},
\mathbf{v}_{j}\right)$. Note that $f(\cdot)$ can be either DNN, inner product,
or other scoring functions.  
For instance, \cite{CKE} proposed CKE, which uses TransE to unify the KG information in the CF framework.
Deep Knowledge-based Network (DKN) \cite{DKN} treats entity embeddings and word embeddings as different channels, where Kim CNN \cite{kimCNN} is used to combine them together for news recommendation.
\cite{KSR} uses a GRU network with a knowledge-enhanced key-value memory network, named KSR, to capture user preferences from the sequential interactions.
\cite{kgat} and \cite{kgin} employ the message passing mechanism of graph neural networks over KG to model the higher-order connections among users, entities, and items.

In the line of path-based methods, recommendations are made based on the connectivity patterns of the entity in user-item graph.
For example, \cite{mcrec} uses CNN to obtain each path instance's embedding.
\cite{kprn} proposed a knowledge-aware path recurrent network (KPRN) solution,
which constructs the extracted path sequence with both the entity embedding and
the relation embedding, and encodes the path sequence with an LSTM layer before
recommendation.  
RippleNet~\cite{ripplenet} combined embedding-based and path-based methods and
modeled users' preference propagation by their historical interests along the
path in the KG. 
Inspired by KG reasoning techniques~\cite{deeppath, multihop, gowalk}, PGPR~\cite{pgpr} performs reinforcement policy-guided path reasoning over
KG-based user-item interaction.
Inspired by such RL-based policy networks, CPR~\cite{RLconversa} models conversational recommendation as a path reasoning problem on a heterogeneous graph.
Different from PGPR which relies on sparse reward signal, ADAC~\cite{ADAC} adopts an Adversarial Actor-Critic model to achieve faster converge. 

These KG reasoning methods (\eg,~\cite{pgpr, ADAC, RLconversa}) can provide explainable reasoning paths while obtaining accurate recommendation results.
However, they do not use temporal information and may lead to inappropriate
recommendations.
Therefore, we propose \tkgrec, aiming to leverage temporal information for promoting explainable recommendation.


\subsection{Temporal Knowledge Graph Embedding}

Existing studies on knowledge graph embedding research focus on static
knowledge graphs. A series of these models are variations of 
TransE~\cite{transE, transD,
transH, transR}. These models assume that facts are not changing with time, which is
obviously contrary to reality. 
Therefore, recent studies begin to take temporal information (\eg , fact
occurrence time, end time, \etc)
%
%
into consideration \cite{surveyKG}, which turns out to further improve the
performance of KG embedding.

There exist only a few studies on temporal knowledge graph (TKG) embedding. 
\cite{leblay} proposed TTransE by simply extending existing
embedding methods like TransE. TTransE adopts translational distance score
functions to learn associations between facts of a KG.
%
\cite{ma_embedding} generalized existing models for static
knowledge graphs to temporal knowledge graphs with a timestamp embedding. 
\cite{hyte} took a timestamp as a hyper-plane and projected
entities and relations to learn KG embedding. 

TKG embedding has to deal with two types of dynamics, namely,
entity dynamics and relation dynamics.

Entity dynamics refers to the fact that real-world events can
change the state of a given entity's state and therefore affect its
corresponding relations. 
\cite{knowevolve} modeled the occurrence of a fact as a
temporal point process, and used a novel recurrent network to learn the
representation of non-linearly evolving entities. 
\cite{contextTemporal} formulated the temporal scoping
problem as a state change detection problem and utilized the context to learn
the states and the state change vectors.

Relation dynamics refers to the fact that the relations
among entities in a knowledge graph may change over time. 
\cite{relation_dy} realized that there exist temporal
dependencies in relational chains following the timeline, for example,
$(P,\text{wasBornIn}, \_ )$ $ \rightarrow$ $ (P, $ $ \text{graduateFrom}$ $, $ $\_ ) $ $\rightarrow $ $(P, \text{workAt}, \_ ) $
$\rightarrow $ $(P,\text{diedIn}, \_ )$, and found that many facts are only
valid during a short time period. To solve that issue, the authors proposed two
time-aware KG completion models to incorporate the above two kinds of temporal
information.

Although those TKG embedding methods achieve better
performance than traditional KG embedding methods, 
none of them have considered if only part of the facts is updated very frequently, which is common in recommendation
and dealed by our \tkgrec.

\subsection{Temporal Information in Recommendation}

Temporal Information is important contextual information for providing users 
with an accurate prediction based on historical behaviors\cite{timeSurveyReview}.
\cite{yi2014beyond} finds that dwell time is an important factor to quantify how likely an item is relevant to a user.
\cite{liang2012time} uses the temporal information in micro-blogs in order to profile users' time-sensitive topic interests.
\cite{chang2017streaming} proposes sREC, which captures temporal dynamics of users and topics under steaming settings, to perform real-time recommendations.
\cite{seqInMusic} proposes CoSeRNN and leverages time context and device context for better music recommendation.
As for temporal patterns of user behaviors, \cite{timeMatters} and \cite{timeAwareSeq} discover ``absolute" and ``relative" time patterns and jointly learn those temporal patterns to model user dynamic preferences and predict future items.
\cite{heteTemporal} defines the periodic and evolving patterns of user preferences, and proposes a cascade of two encoders with an attention module to capture such temporal patterns.

Recently, KG-based methods achieve great success for many temporal tasks.
\cite{kerl} uses both sequence- and knowledge-level state representations to capture user preference with an induction network, and adopt a composite reward to capture both sequence-level and knowledge-level information for sequential recommendation.
\cite{KSR} uses GRU to capture the user’s sequential preference, and utilizes knowledge base information to model the user’s attribute-level preference.
\cite{huang2019explainable} designs EIUM, which represents KG with 
multi-modal fusion, and adopts a masked self-attention model to encode the user’s sequential interactions for capturing her dynamic interests.
Most of those methods with temporal information consider KG information as enhanced representation to assist sequential recommendation.
Our work differs from these works by modeling the statistical and structural 
temporal patterns between user-item interactions to construct TCKG, and propose time-aware path reasoning over the TCKG to make full use of the structural constraints in KG. It's inconvenient to utilize our \tkgrec\ to predict the next item task like other sequential recommendation methods, because the KG data is different from the sequence data. That is, we need to construct a dynamic edge MASK matrix on KG's adjacency matrix to avoid the next item $v$'s information and other items behind $v$ from influencing the training phase. We will leave this as our future work.

%% file: 6-Conclusion.tex
\section{Conclusion}
\label{sec:conclusion}

In this paper, we present \tkgrec, a novel 
time-aware path reasoning model for recommendation, which addresses the timeliness of
interpretation.
To the best of our knowledge, the \tkgrec\ method is the first work to
introduce time-aware path reasoning method into a recommendation
system and achieves significant performance improvement by leveraging the
temporal information.
%
%

Our model first uses time-aware interaction relation extraction component to
construct TCKGs for the
reasoning environment. Then the time-aware representation learning component encodes 
entities and relations to get proper
representation. Finally is the key time-aware reasoning component, which adopts a 
time-aware
personalized reasoner to explore temporal information assisted recommendation.
We also conduct extensive experiments on three real-world datasets to evaluate
the effectiveness of \tkgrec. Evaluation results demonstrate that \tkgrec\
outperforms the existing models on a set of widely used metrics.

For future work, we plan to evaluate \tkgrec\ using more datasets from other
product domains and other major online vendors, transform the training process 
to make it suitable for next item prediction,
and extend our \tkgrec\ model
by leveraging the adversarial learning model to automatically shape rewards in
order to achieve more accurate recommendation results.  We also plan to combine
causal inferences with our model to achieve better interpretability.

%% file: acknow.tex
\section{ACKNOWLEDGMENTS}
This work is supported in part by National Key R\&D Program of China (Grant No. SQ2021YFC3300088), the National Natural Science Foundation
of China (Grant 62102382, U19A2079, No. U19B2036),  USTC Research
Funds of the Double First-Class Initiative (WK2100000019), and the National Key R\&D Program of China (2021ZD011802).